\documentclass{aa}  

\usepackage{graphicx}
\usepackage{txfonts}
\usepackage{breqn}
\usepackage{subcaption}

\usepackage{array}
\usepackage{rotating}
\usepackage{multirow}

\providecommand{\tabularnewline}{\\}

\newcommand{\specialcell}[2][c]{%
  \begin{tabular}[#1]{@{}c@{}}#2\end{tabular}}

\begin{document}

   \title{PELICAN: deeP architecturE for the LIght Curve ANalysis  }


\author{Johanna Pasquet\inst{1}, Jérôme Pasquet \inst{2}, Marc Chaumont \inst{3} and Dominique Fouchez\inst{1}}

\institute{
	\inst{1} Aix Marseille Univ, CNRS/IN2P3, CPPM, Marseille, France \\
    \inst{2} AMIS, Université Paul Valéry, Montpellier, France \\ \qquad TETIS, Univ. Montpellier, AgroParisTech, Cirad, CNRS, Irstea, Montpellier, France \\ \qquad Aix-Marseille Universit\'e, CNRS, ENSAM, Universit\'e De Toulon, LIS UMR 7020\\
    \inst{3} LIRMM, Univ. Montpellier, CNRS, Univ. Nîmes, France
 }

   \date{}

 
  \abstract{
We developed a deeP architecturE for the LIght Curve ANalysis (PELICAN) for the characterization and the classification of light curves. It takes light curves as input, without any additional features. PELICAN can deal with the sparsity and the irregular sampling of light curves. It is designed to remove the problem of non-representativeness between the training and test databases coming from the limitations of the spectroscopic follow-up.  We applied our methodology on different supernovae light curve databases.
First, we evaluated PELICAN on the Supernova Photometric Classification Challenge for which we obtained the best performance ever achieved with a non-representative training database, by reaching an accuracy of 0.811. Then we tested PELICAN on simulated light curves of the LSST Deep Fields for which PELICAN is able to detect 87.4\% of supernovae Ia with a precision higher than 98\%, by considering a non-representative training database of 2k light curves. PELICAN can be trained on light curves of LSST Deep Fields to classify light curves of LSST main survey, that have a lower sampling rate and are more noisy. In this scenario, it reaches an accuracy of 96.5\% with a training database of 2k light curves of the Deep Fields.
It constitutes a pivotal result as type Ia supernovae candidates from the main survey might then be used to increase the statistics without additional spectroscopic follow-up. Finally we evaluated PELICAN on real data from the Sloan Digital Sky Survey. PELICAN reaches an accuracy of 86.8\% with a training database composed of simulated data and a fraction of 10\% of real data.
The ability of PELICAN to deal with the different causes of non-representativeness between the training and test databases, and its robustness against survey properties and observational conditions, put it on the forefront of the light curves classification tools for the LSST era.}

   \keywords{methods: data analysis --
                techniques: photometric --
                supernovae: general
               }
\titlerunning{PELICAN}
\authorrunning{J. Pasquet et al.}

   \maketitle
%

\section{Introduction}
A major challenge in cosmology is to understand the observed acceleration of the expansion of the universe. A direct and very powerful method to measure this acceleration is to use a class of objects, called standard candles due to their constant intrinsic brightness, which are used to measure luminosity distances. Type Ia supernovae (SNe Ia), a violent endpoint of stellar evolution, is a very good example of such a class of objects as they are considered as standardizable candles. The acceleration of the expansion of the universe was derived from observations of several tens of such supernovae at low and high redshift \citep{Perlmutter1999, Riess1998}. Then, several dedicated SN Ia surveys have together measured light curves for over a thousand SNe Ia, confirming the evidence for acceleration expansion \citep[e.g.][]{Betoule2014,Scolnic2018}. \\
The future Large Survey Synoptic Telescope \citep[LSST,][]{LSST2009} will improve on past surveys by observing a much higher number of supernovae. By increasing statistics by at least an order of magnitude and controlling systematic errors, it will be possible to pave the way for advances in precision cosmology with supernovae.\\
A key element for such analysis is the identification of type Ia supernova. But the spectroscopic follow-up will be limited and LSST will discover more supernovae than can be spectroscopically confirmed. Therefore an effective automatic classification tool, based on photometric information, has to be developed to distinguish between the different types of supernovae with a minimum contamination rate to avoid bias in the cosmology study. 
This issue was raised before and has led to the launch of the Supernova Photometric Classification Challenge in 2010 (SPCC, \citeauthor{Kessler2010a}) to the astrophysical community. Several classification algorithms were proposed with different techniques resulting in similar performance without resolving the problem of non-representativeness between the training and test databases. Nonetheless, the method developed by Sako et al. (\citeyear{Sako2008}, \citeyear{2018PASP..130f4002S}) based on template fitting, shows the highest average figure of merit on a representative training database, with an efficiency of 0.96 and an SN Ia purity of 0.79. \\
Since then, several machine learning methods were applied to classify supernovae light curves \citep[e.g. ][]{Richards2012, Ishida2013, Karpenka2013, Varughese2015, Moller2016, Lochner2016, Dai2018}. They showed interesting results when they are applied on a representative training dataset but the performance dramatically decreases when the learning stage is made on a non-representative training subset, which represents however the real scenario.\\
We propose to explore in this paper a new branch of machine learning, called deep learning, proved to be very efficient for image and time series classification \citep[e.g. ][]{googlenet,resnet,time1}. One of the main difference with the classical machine learning methods is that the raw data are directly transmitted to the algorithm that extracts by itself the best feature representation for a given problem. In the field of astrophysics, deep learning methods have shown better results than the state of the art applied to images for the classification of galaxy morphologies \citep{Morpho2018}, the classification of transients \citep{Buisson2015,Gieseke2017} and the estimation of photometric redshifts \citep{Pasquet2018} to name a few. This method have also showed impressive performance for the classification of light curves \citep{Mahabal2017, 2018A&A...611A..97P} and especially the classification of supernovae \citep{Charnock2017,brunel}.

In this work we develop a complex Deep Learning architecture to classify light curves. We apply our study to the classification of light curves of supernovae. Unlike the other studies, our method overcome the problem of non-representativeness between the training and the test databases, while considering a small training database. We apply our method on the SPCC challenge, then on LSST simulated data including a biased and small training database. We also validate our method on real data coming from the Sloan Digital Sky Survey (SDSS) data. The paper is organized as follows. In Section 2, we explain the different issues for the classification of light curves. In section 3, we introduce deep learning concepts that we used and developed in this work. In Section 4, we present our architecture named PELICAN (deeP architecturE for the LIght Curve ANalysis). In section 5, we describe the different datasets used in this study. In Section 6 we present the experimental protocol that we adapted to make PELICAN robust against the differences of sampling and noise. In Section 7 we present our results for different databases. In Section 8 we analyze the behaviour of PELICAN with respect to a number of light curve properties and observational conditions. Finally we conclude and expose perspectives in Section 9.

\section{Light curve classification issues }
\begin{figure}[h!]
\center
\includegraphics[width=4.4cm]{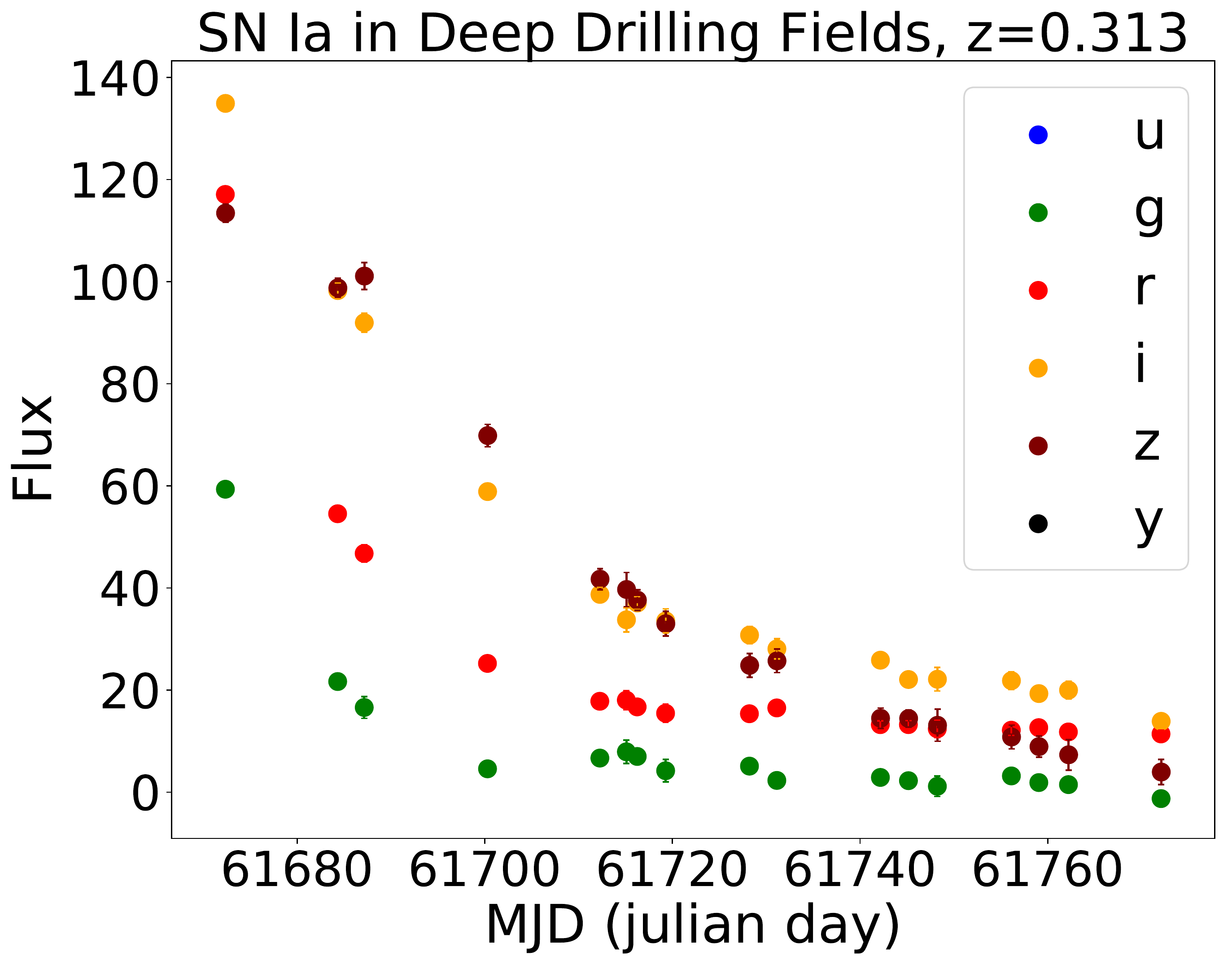}
\includegraphics[width=4.4cm]{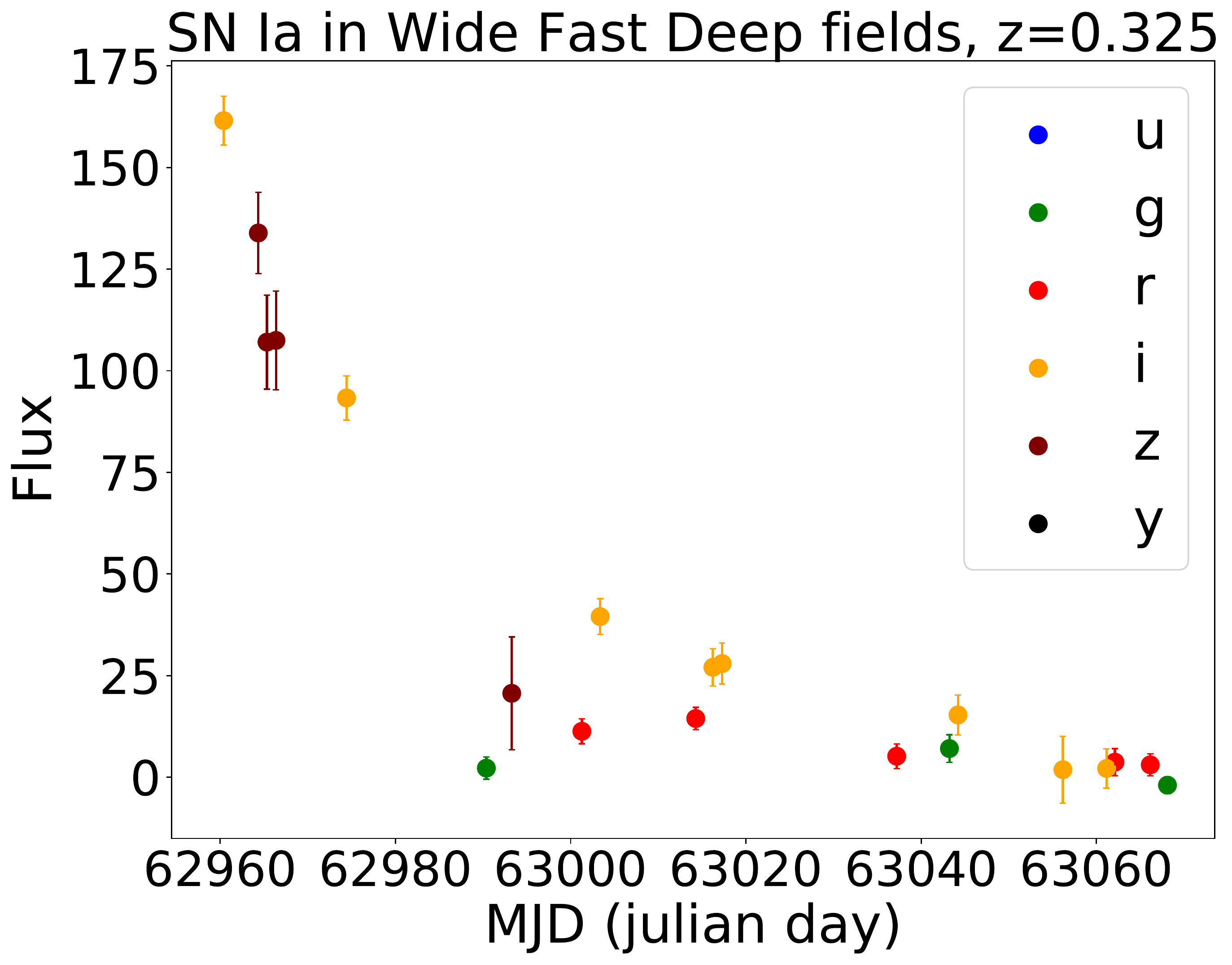}
\caption{Example of two light curves of type Ia supernovae observed for a high LSST cadence (Deep Drilling Fields), on the left and a low LSST cadence (Wide Deep Fast), on the right, at two similar redshifts.}
\label{lc}
\end{figure}

Light curves are fundamental signals to measure the variability of astrophysical objects. They represent the flux of an object along time in different photometric bands (e.g. \textit{ugriz} system). Due to the observational strategy and conditions, light curves have an irregular sampling, often sparse. Therefore a sampling with two different observational cadences present several differences. Figure \ref{lc} shows, as an example, two simulated light curves with two different cadence models (see Section \ref{lsst_simu_sec}). 
Compared to an image, such light curves have incomplete and non-continuous information, thus imposing dedicated training algorithms.

\begin{figure}[h!]
\includegraphics[width=9.0cm]{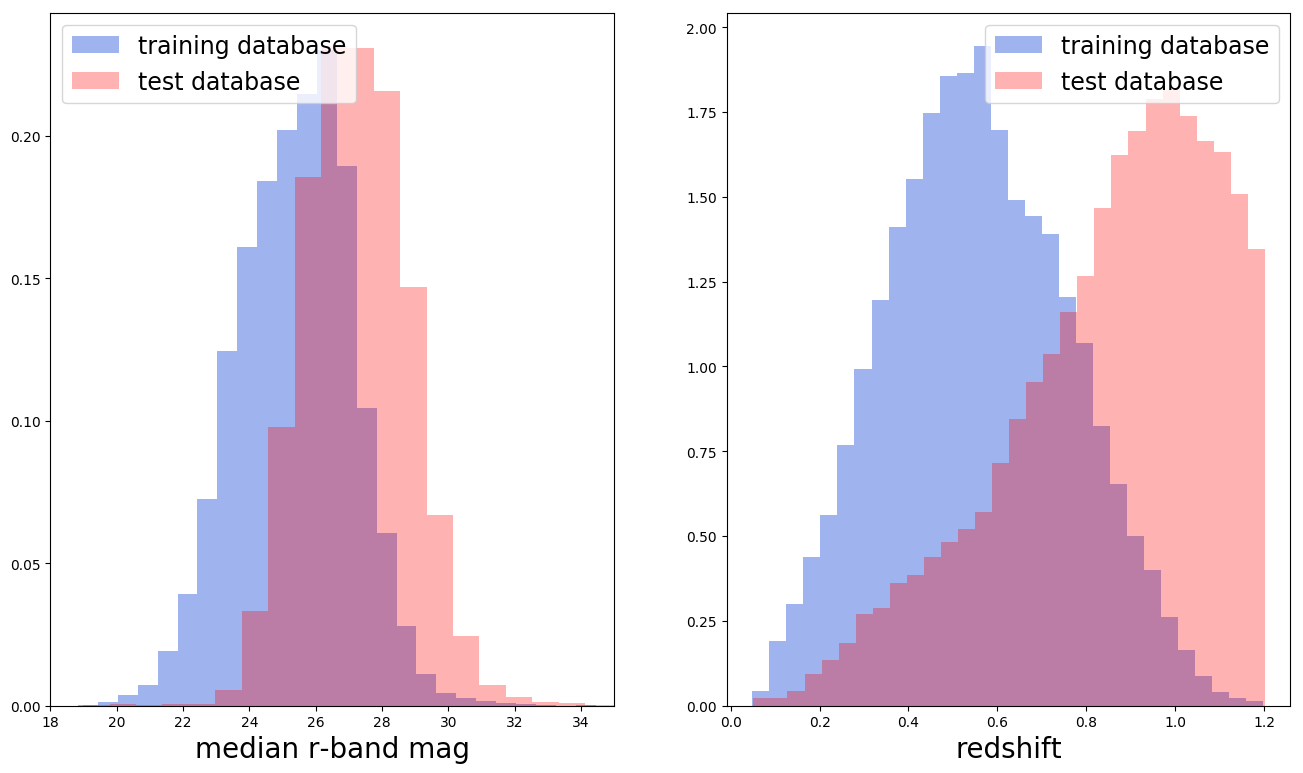}
\caption{Distributions of LSST simulated data of the median r-band magnitude (on left) and the simulated redshift (on right) for the training dataset in blue and the test dataset in red. The mismatch is well visible as there is a significant shift between the two distributions.   }
\label{mismatch}
\end{figure}

The other issue is the non-representativeness between the training and the test databases.
As the spectroscopic follow-up, used to label the lightcurves, is limited, the coverage of the training database in brightness, redshift and number is different from the test database as shown on Fig. \ref{mismatch}. Moreover the absolute magnitude of SN Ia is correlated with two quantities. First, brighter SN Ia have wider, slower declining light curves. This variability can be described as a timescale stretch of the light curve \citep{Phillips1993ApJ...413L.105P}. In addition brighter SN Ia are bluer and a color correction has to be applied to standardize them \citep{van1995ApJ...453L..55V,Tripp1998A&A...331..815T}. So due to these correlations, the lower-stretch and redder supernovae are fainter and tend to have small recovery efficiency (Malmquist bias) and so are under-represented in the training database which is limited in brightness. The non-representativeness of the databases, which is a problem of mismatch, is critical for machine learning process. \\

In general, machine learning methods require a sufficient large number of training data in order to correctly classify, so the small size of the training database involves another difficulty.

To provide a solution for each of these issues, we have designed a specific architecture. First, light curves from the test database are trained with a non-supervised model without using the knowledge of labels. This allows to reduce the mismatch between the training and the test databases and provides a solution to the small training dataset, by extracting features from the larger test database. \\
To reduce again the problem of non-representativeness we performed a second training step to minimize the distances in the feature representation space between bright and faint supernovae of same labels and maximize distances of supernovae with different labels. \\
Finally we integrated a regularization term into the training to adapt the model to the sparsity of data. \\
The resulting deep architecture, dedicated to the characterization and classification of light curves, is presented in Section \ref{ourArchi}.

\section{Deep Learning Model}
In this section, we present the main deep learning concepts which are used to build our network architecture. Namely the convolution layer network which describes the basics, the autoencoder which is a non-supervised module where we detail the notion of loss function and finally the contrastive approach where an adapted loss function is defined to improve the clustering of same label entries.

\subsection{Convolutional Neural Network}
\label{cnn}
Convolutional Neural Network (CNN) is a special type of multilayered neural network that is made up of neurons that have learnable weights and biases. The architecture of a CNN is designed to take advantage of the 2D structure of an input image. It takes as input a $h\times w \times c$ image where $h$ is the height and $w$ is the width of the image and $c$ is the number of channels. As a light curve is a 1D signal, we transformed it into a 2D "light curve image" (LCI) as we did in \cite{2018A&A...611A..97P}. The width of the LCI is the temporal axis (expressed in days) and the height is the number of photometric bands. For example, if we consider a light curve measured in \textit{ugriz} bands in 200 days, the corresponding LCI is a 2D array of dimension ($5\times 200$) pixels. As a light curve is not a continuous signal, the corresponding array is composed of many blank cells that we filled with zero values. A harmful consequence is the overfitting of the position of missing data, which could dramatically degrade the performance. In this case, the model learns the exact position of missing data of the training light curves and is not able to generalize to the unseen test data. To prevent this overfitting and make the model invariant against the position of missing data we have integrated several techniques that are explained in Section \ref{exp}.

\subsubsection{Convolution layers}
In a convolution layer, each neuron applies a convolution operation to the input data using a 2D map of weights used as a kernel. Then resulting convolved images are summed, a bias is added and a non-linearity function is applied to form a new image called a feature map. In the first convolution layer, the convolution operation is realized between the input LCI and the set of convolution kernels to form feature maps that are then convolved with convolution kernels in the next convolution layer. 
For the non-linearity function, we mainly use the most commonly used activation function :  the ReLU \citep[Rectified Linear Unit,][]{ReLU} defined by $f(x)= max(x, 0)$.  
 The weights of the kernels are updated during the training by back-propagation process. 

\subsubsection{Pooling layers}
The network can be composed of pooling layers which quantify the information while reducing the data volume. The two most used methods consist in selecting only the maximum or the average value of the data in a local region. \\

\subsubsection{Fully connected layers}
Finally, fully connected layers are composed of neurons that are connected to every neuron of the previous layer and perform the classification process with features from previous convolution layers. \\
More details on CNN models can be found in \cite{2018A&A...611A..97P,Pasquet2018}.

\subsection{Autoencoder}
\label{auto_explication}
\begin{figure}[h!]
\center
\includegraphics[width=9.2cm]{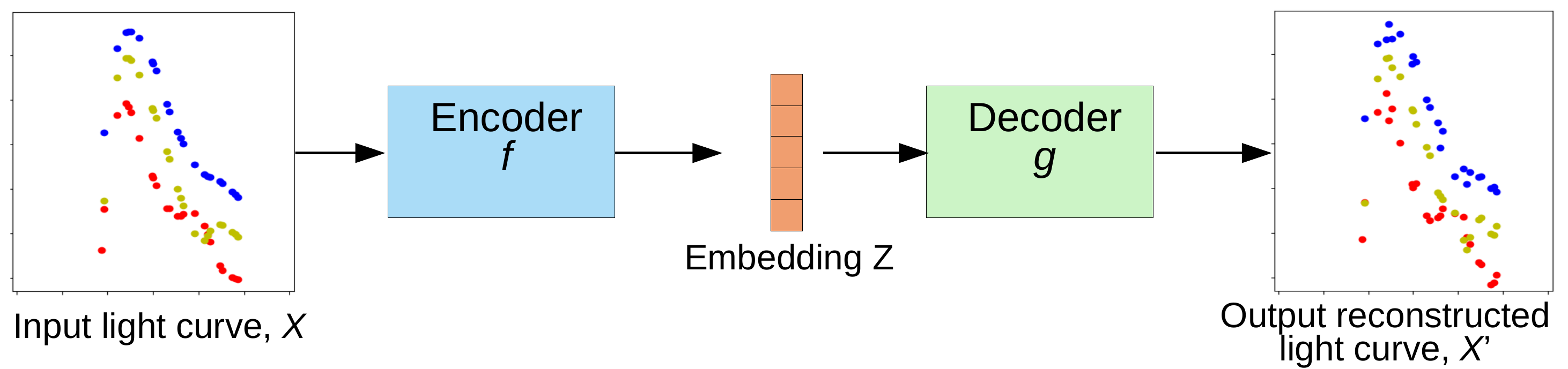}
\caption{Schema of the autoencoder process.}
\label{autoen}
\end{figure}

To benefit from the information of the light curves of the test database and so reduce the mismatch between the training and test databases, we have adapted a non-supervised autoencoder.
An autoencoder is an unsupervised learning algorithm that tries to learn an approximation to the identity function such as output should mimic the input. As a consequence the internal layers exhibit a good representation of the input data. The input $X \in \mathbb{R}^{D}$, with $D=h\times w$ is transformed into an embedding $Z \in \mathbb{R}^{K}$, often such as $K<<D$. The mapping from $X$ to $Z$ is made by the encoder, noted $f$, that could perform a dimension reduction to finally get a good representation of the data in a compressed format. The reconstruction of the original signal, $X' \in \mathbb{R}^{D}$ , is obtained by the decoder, noted $g$, that uses the compressed embedding representation $Z$ (see Fig. \ref{autoen}). 
The objective of the autoencoder is to minimize the distance function (for example L2 distance), named loss function, between each input $X$ and each output $X'$. The learning process of the autoencoder consists in iteratively refining its internal parameters such that the evaluation of the loss function, on all the learning set, is reduced. 
The loss function associated to the autoencoder, noted  $L_{auto}$ is defined as follow: 

\begin{equation}	
L_{auto} = ||X-g(f(X))||,
\label{autoEq}
\end{equation}
with $X$ represents the input signal and ||.|| symbolizes the L2 distance.

The minimization of the loss function that maps from $X$ to $Z$ in order to obtain $X'$ does not guarantee the extraction of useful features. Indeed the network can achieve a perfect reconstruction by simply "copying" the input data and thus obtain a minimal mapping error. Without any other constraints, the network can miss a good representation of the input data. A strategy to avoid this problem is to constrain the reconstruction criterion by cleaning or denoising partially corrupted input data with a denoising autoencoder \citep{autoencoderDenoise}. 

\begin{figure}[h!]
\center
\includegraphics[width=9.2cm]{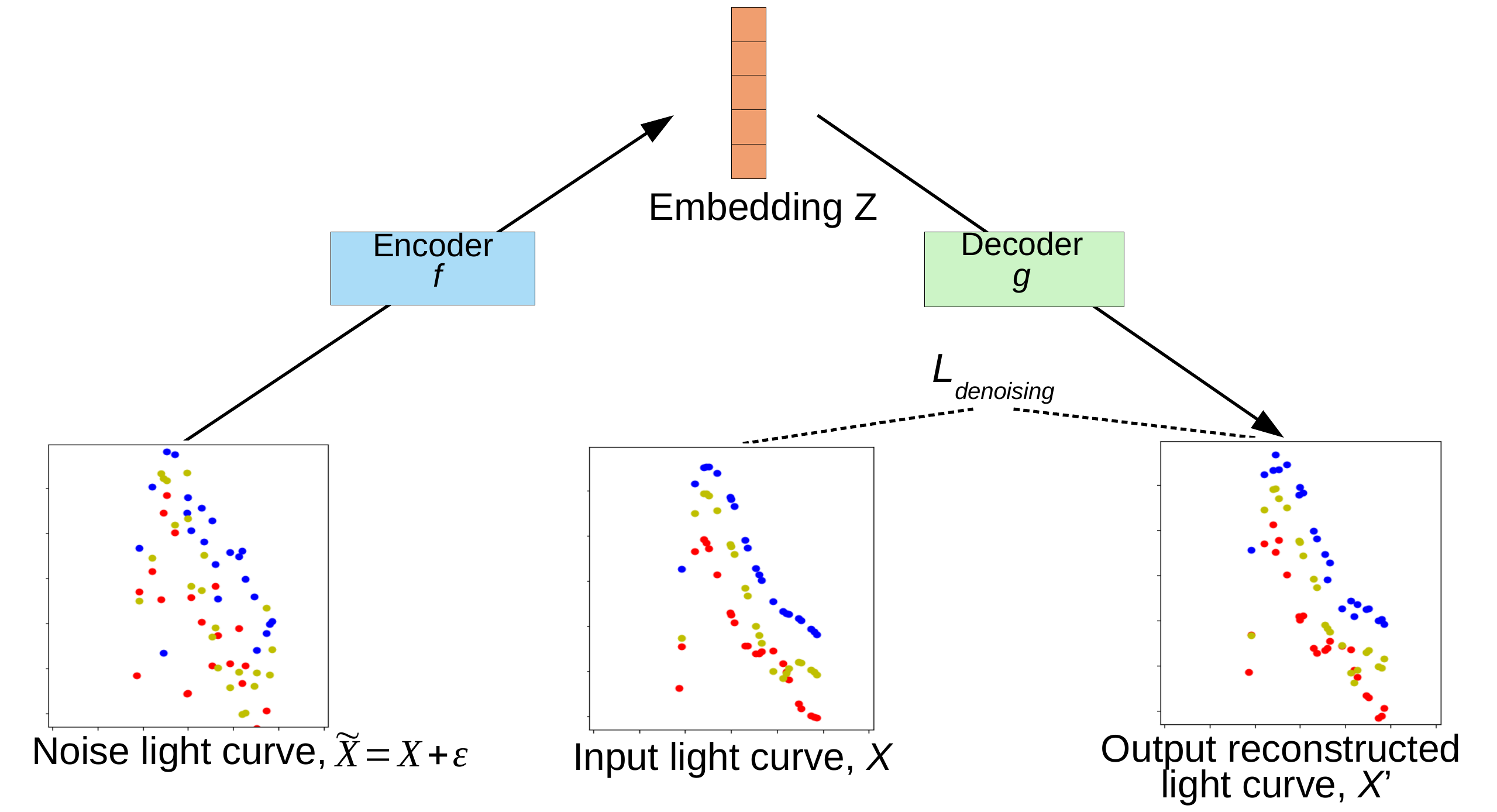}
\caption{Schema of the denoising autoencoder process.}
\label{denoise}
\end{figure}

The methodology consists to first apply noise, for example an additive Gaussian noise, on input data to corrupt the initial input $X$ into $\widetilde{X}$. Then the autoencoder maps $\widetilde{X}$ to $Z$ via the encoder $f$ and attempt to reconstruct $X$ via the decoder $g$ (see Fig. \ref{denoise}). Although $Z$ is now obtained by applying the encoder on corrupted data $\widetilde{X}$, the autoencoder is still minimizing the reconstruction loss between a clean $X$ and its reconstruction from $\widetilde{X}$ with the following loss function:

\begin{equation}
L_{denoising} = ||X-g(f(\widetilde{X}))||,
\end{equation}
where $\widetilde{X}=X+\epsilon$, with $\epsilon$ an additive uniform noise with the same dimension of $X$ (see Fig. \ref{denoise}).

As the number of parameters of the autoencoder is high, light curves are sparse and the size of the training database that we will have is small, the overfitting can seriously affect the performance of the network in our case. A solution is to introduce sparsity to the learned weights to avoid learning "noisy" patterns. The sparsity concept does the assumption that only a few neurons are required in order to reconstruct a complete signal.
Let us consider that a neuron is active if its output value is close to 1 and inactive if its output value is close to 0. We want a very sparse representation such that the neurons of a given layer should be inactive most of the time and so obtain an average activation of neurons close to 0. Therefore the output signal is reconstructed with a very limited number of activated neurons.

We note $\hat{\rho_{j}}$ the average activation of hidden unit $j$ over the training set. To force neurons to be inactive, we enforce the constraint: $\hat{\rho_{j}} = \rho$, with $\rho$ a \textit{sparsity parameter} that is initialized close to zero. Thus the average activation of each hidden neuron has to be close to a small value and so most of the neurons have to be inactive to satisfy this constrain.

In practice, a penalty term is added to the loss function to penalize $\hat{\rho_{j}} = \rho$ deviating significantly from $\rho$. One of the main frequently used penalty term is the Kullback-Leibler (KL) divergence \citep{Hinton:2002:TPE:639729.639730} defined as:

\begin{equation}
KL(\rho ||\hat{\rho_{j}})  =  \rho \,  \textrm{log}\frac{\rho}{\hat{\rho_{j}}} + (1-\rho)\,\textrm{log} \, \left(\frac{1-\rho}{1-\hat{\rho_{j}}}\right).
\end{equation}

KL divergence measure how two distributions are different from one another. It has the property that $KL(\rho ||\hat{\rho_{j}}) = 0 $ if $\hat{\rho_{j}} = \rho$ and otherwise it increases monotonically as $\hat{\rho_{j}}$ diverges from $\rho$.

The loss function is now integrating this penalty term as:

\begin{equation}
L_{sparse} = ||X-g(f(X))|| + \alpha \Omega_L ,
\label{lsparse}
\end{equation}

with $\alpha \in \mathbb{R}$ a scalar that weights the sparse regularization term $\Omega_L$ which is the sum of the KL divergence term for neurons of a chosen layer numbered $L$ defined as:
\begin{equation}
\Omega_L =  \sum_j KL(\rho||\hat{\rho_{j}}) = \sum_j \rho \textrm{log}  \frac{\rho}{\hat{\rho_{j}}}  + (1-\rho) \textrm{log}  \frac{1-\rho}{1-\hat{\rho_{j}}}, 
\label{omega}
\end{equation}
$j \in \{1,...,N_{j}\}$ with $N_{j}$ the number of neurons of layer $L$.

\subsection{The contrastive loss function}

The contrastive loss function was introduced to perform a dimensionality reduction by ensuring that semantically similar examples are embedded close together \citep{contrastive}. It was shown that this method provides invariance to certain transformations on images. The contrastive loss function is computed over pairs of samples unlike traditional loss functions which is a sum over all the training database. 
Let ($\vec{Z_{1}}$, $\vec{Z_{2}}$) a pair of input data and $Y$ a binary label assigned to this pair. If $\vec{Z_{1}}$ and $\vec{Z_{2}}$ have the same label, $Y=0$, otherwise $Y=1$. The distance function between $\vec{Z_{1}}$ and $\vec{Z_{2}}$ is learned as the euclidean distance: $D=||\vec{Z_{1}} - \vec{Z_{2}}||$. Thus the loss function tends to maximize the distance $D$ if they have dissimilar labels and minimize $D$ if they have similar labels. So we can write the loss function as:
\begin{equation}
L(D,(Y, \vec{Z_{1}},\vec{Z_{2}})) = (1-Y)L_{S}(D) + Y\,L_{D}(D)
\end{equation}
with $(Y, \vec{Z_{1}},\vec{Z_{2}})$ a labeled sample pair and $L_{S}$ the partial loss function for a pair of similar labels, $L_{D}$ the partial loss function for a pair of dissimilar labels. To get low values of $D$ for similar pairs and high values of D for dissimilar pairs, $L_{S}$ and $L_{D}$ must be designed to minimize $L$.
We introduce the margin $m>0$ which defines a minimum distance between $(\vec{Z_{1}},\vec{Z_{2}})$. Dissimilar pairs contribute to the loss function only if their distance is below this minimum distance so that pairs who share a same label will be bring closer, and those who does not share a same label will be drive away if their distance is less than $m$. The final contrastive loss function is defined as:

\begin{equation}
L_{contrastive} = (1-Y) \, ||\vec{Z_1}-\vec{Z_2}||+Y \, \textrm{max} \left(0, m-\sqrt{||\vec{Z_1}-\vec{Z_2}||}\right)^2
\label{contra}
\end{equation}

\section{The proposed architecture}
\label{ourArchi}
We developed the method named \textit{deeP architecturE for the LIght Curve ANalysis} (PELICAN) to obtain the best feature-space representation from light curves and perform a classification task. In this work, we apply PELICAN for the classification of supernovae light curves but it can be extended to the classification of other variable or transient astrophysical objects. \\ 
PELICAN is composed of three successive modules (see Fig. \ref{netork}). Each of them has a specific purpose with a loss function associated. 
The first module learns a deep representation of light curves from the test database under an unsupervised autoencoder method. 
The second module optimizes a contrastive loss function to learn invariance features between the bright and fainter supernovae from the training database.
Finally, the third module performs the classification task

In this section we explain in more details the different mechanisms and objectives of the operations related to each module.

\subsection{The autoencoder branch}
To deal with the low number of examples in the training database that leads to overfitting and mismatch between the spectroscopic and photometric distributions (see Fig. \ref{mismatch}), we propose to train an unsupervised sparse autoencoder method on the test database. In this way we can benefit from information of light curves in the test database without knowing the label associated to each object. 

\begin{figure*}
\center
\includegraphics[width=18cm]{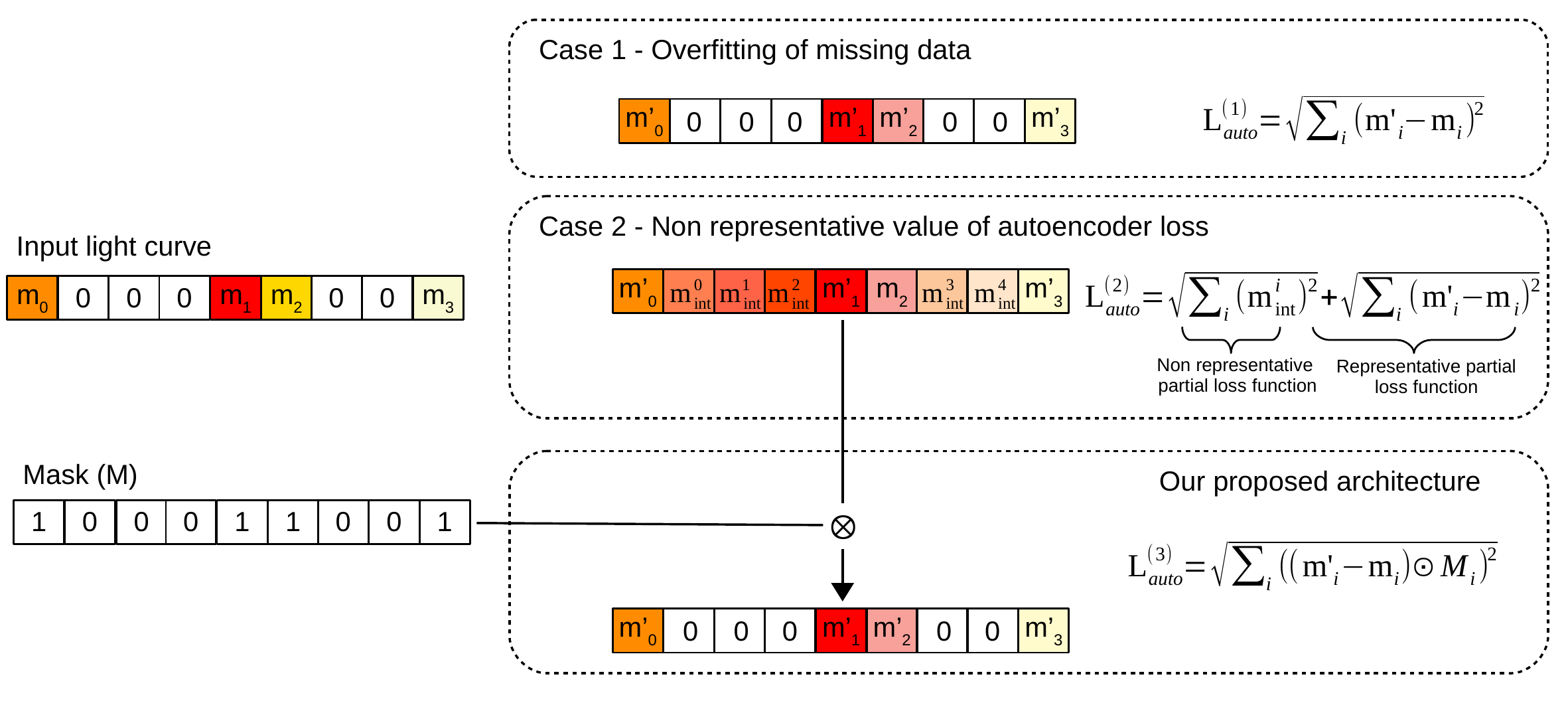}
\caption{Illustration of the overfitting of the missing data that could appear in the autoencoder process and the solution proposed to overcome it. The input light curve is composed of different magnitudes ($m_0,\,m_1,\,m_2\,m_3$) and missing values represented by zero values. In case 1, the algorithm has completely overfitted the missing data by replacing them at the same position on the light curve. So the loss function, $L_{auto}^{(1)}$ is ideally low. In case 2 the algorithm has completed the missing data by interpolated them. However as the computation of the loss is made between the new values of magnitudes, ($m_{int}^{0},\,m_{int}^{1},\,m_{int}^{2},\,m_{int}^{3},\,m_{int}^{4}$), compared to zero values, the value of the loss $L_{auto}^{(2)}$ is overestimated. The solution that we provided is to multiply the interpolated light curve by a mask $M$ before the computation of the loss, $L_{auto}^{(3)}$.}
\label{autopro}
\end{figure*}

The autoencoder takes as input a batch of LCIs of size $h\times w$ from the test database, that are encoded and decoded through a convolutional neural network architecture (CNN). To extract useful features, we applied an uniform noise, which affects differently each magnitude on the light curve by adding a random value $\in [-0.5, 0.5] \,\textrm{mag}$, before passing through the encoder (see Fig. \ref{denoise}).  \\ 
In the first part of the CNN, the encoder which is composed of 9 convolution layers (conv 1 to conv 9 in Fig. \ref{netork}) and 4 pooling layers (Pool 1,4, 6 and 8), converts the input noisy LCIs into an embedding representation. Then, the decoder reconstructs the original LCIs from the embedding representation through two fully connected layers (FC10 and FC11) of 5000 neurons. So the output of the autoencoder is a reconstructed LCI with the same size than input, $h\times w$. The loss function associated to the autoencoder, called Autoencoder Loss on Fig. \ref{netork}, minimizes the difference between the original LCI and the reconstruction one. However, we have to pay attention at the overfitting of missing data on the light curve. 
An illustration of the overfitting problem and the solution we propose is given on Fig. \ref{autopro}. 
By construction, the input LCI is composed of many zero values (see Section \ref{cnn}), and the autoencoder model can "learn" the position of missing data and just replace them at the same position on the reconstructed LCI (see case 1 on Fig. \ref{autopro}). To avoid this overfitting, we introduce a regularization term. In this case, the autoencoder learns to interpolate zero values (case 2 on Fig. \ref{autopro}). But the computation of the autoencoder loss can not take into account these interpolated values as they are compared to zero values on the initial LCI, and so lead to a divergence of the loss function. Therefore we propose to define a mask with the same size as the considered original light curve, filling with 1 if there is an observation on the light curve, and 0 otherwise. The reconstructed LCI is then multiplied by the mask before the minimization of the loss function (case 3 on Fig. \ref{autopro}). Equation \ref{autoEq} becomes:
\begin{equation}	
L_{auto} = ||X-g(f(X)) \odot M(X)||
\label{}
\end{equation}
with $M(X)$ the mask related to the input light curve $X$.\\
Finally, we compute the penalty term as defined in equation \ref{omega}, in the second fully connected layer, FC11 and called it Sparsity loss. It depends on two hyperparameters: the sparsity parameter $\rho$ and the weight of the sparse regularization $\alpha$. To determine the best values of $\rho$ and $\alpha$, we searched the best combination using a 2D grid search among values into the following finite sets:  $\{10^{-5}, 5 \times 10^{-4},  5 \times 10^{-3}, 10^{-3}, 5 \times 10^{-2}, 10^{-2},  5\times10^{-1}, 10^{-1}\}$ and $\{10^{-3}, 5 \times 10^{-2}, 10^{-2},  5 \times 10^{-1}, 10^{-1}\}$ respectively.\\
However the regularization term does not take into account the number of observations on each light curve which varies significantly. It may cause overfitting as the number of active neurons is then always the same whatever the number of data points on each light curve. So the number of active neurons has to be adapted depending on the number of observations in all filters. Thus, we propose to express the sparsity parameter, $\rho$, as a linear function depending on the number of observation for each light curve. This contribution allows to increase the number of active (inactive) neurons when the light curve is densely (poorly) populated with observations.
We define a new sparsity parameter $\rho'(l)$ for the specific light curve noted $l$ as follow:
\begin{equation}
\rho'(l) = \rho_{a} n_{l} + \rho_{b}
\end{equation}
with $n_{l}$ the number of observations on the light curve $l$, $\rho_{a}$ and $\rho_{b}$ are two hyper-parameters. They are determined as the same time as $\alpha$ using a 3D grid search among the same values as $\rho$.

In this case, 
the sparse regularization term (see equation \ref{omega}) of our autoencoder module take the form : 

\begin{equation}
\label{eq_sparse}
\Omega'_L(l) = \sum_j \rho'(l) \textrm{log}  \frac{\rho'(l)}{\hat{\rho_{j}}}  + (1-\rho'(l)) \textrm{log}  \frac{1-\rho'(l)}{1-\hat{\rho_{j}}} 
\end{equation}

\subsection{The contrastive branch}
Once the autoencoder training has converged on the test database the weights of its convolution and fully connected layers are fixed. Then, the output of a chosen layer of the encoder part is given as input to the contrastive branch. This second module is designed to reduce the mismatch between the training (higher magnitudes) and the test (lower magnitudes) databases. This requires a specific contrastive loss function that is minimized through a CNN architecture. So we propose a loss function that minimizes the variations of intra-class light curves and maximizes the variations of inter-class light curves. In this way, we split the training database in four subsets following a cut magnitude $m^{c}$ in the i-band magnitude.


If we note $m^{Ia} (l)$ the i-band median magnitude of a type Ia light curve and $m^{non-Ia} (l)$ the i-band median magnitude of a non-Ia type light curve, a given light curve can belong to one of the four following subsets:
\begin{itemize}
\item LC1 : type Ia light curves with  $m^{Ia} (l) < m^{c} $,
\item LC2 : type Ia light curves with  $m^{Ia} (l) > m^{c}$,
\item LC3 : non-Ia type light curves with  $ m^{non-Ia} (l) < m^{c}$,
\item LC4 : non-Ia type light curves with $ m^{non-Ia} (l) >m^{c}$
\end{itemize}
Therefore the goal is to define a loss function that minimizes the variation between intra-class light curves, i.e between the {LC1-LC2} and {LC3-LC4} sets; and maximizes the variation between inter-class light curves, i.e between {LC1-LC3}, {LC1-LC4}, {LC2-LC3}, and {LC2-LC4} sets.

\begin{figure}[h!]
\includegraphics[width=\textwidth / 2]{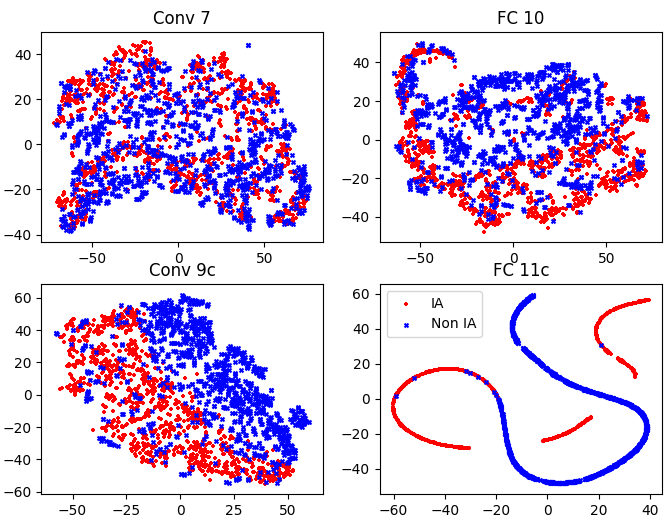}
\caption{The t-distributed stochastic neighbor embedding (t-SNE) projections with features extracted from two layers of the autoencoder module (Conv 7 and FC 10) and from two layers of the contrastive module (Conv 9c and FC 11c).}
\label{tsne_contrastive}
\end{figure}

Equation \ref{contra} becomes:

\begin{small}
\begin{align}
\label{contrast_eq}
L= &\frac{1}{2} \textrm{max} \left( 0, m-\sqrt{||LC1-LC3||} \right)^2 + \frac{1}{2} \textrm{max} \left(0, m-\sqrt{||LC1-LC4||}\right)^2 + \nonumber \\ & 
\nonumber \frac{1}{2}\textrm{max} \left(0, m-\sqrt{||LC2-LC3||}\right)^2 + \frac{1}{2}\textrm{max}\left(0, m-\sqrt{||LC2-LC4||}\right)^2 + \\  & 
||LC1-LC2|| + ||LC3-LC4||
\end{align}
\end{small}
We introduce $\frac{1}{2}$ terms into the formula to weight the inter-class distances so that the inter-class and the intra-class distances have the same weight in the computation of the loss function. \\
 In practice this means that the encoder is fed with sets of four light curves from the training database, with one light curve from each subset. The learning of the contrastive branch (see Fig. \ref{netork}) is done without updating the training weights of the autoencoder, that have been adjusted during the non-supervised step on the test database. This step allows also to solve the problem of asymmetry that exists between the classes as this module takes as input both light curves of type Ia and non-Ia supernovae at the same time. Features from the seventh convolution (conv 7 on Fig. \ref{netork}) are given as input to the contrastive branch where the training weights are updated. Therefore the minimization of the contrastive loss is made only on the training database. The choice of the seventh convolution layer as input to the contrastive branch was made for several reasons. First of all, as the encoder part of the first module is dedicated to extract relevant features from the test light curves to characterize them precisely, while the decoder part is designed to reconstruct the original light curve, we decided to extract features from the first part of the autoencoder to reduce the mismatch between the training and the test databases. Figure \ref{tsne_contrastive}, which represents the t-SNE\footnote{The t-distributed stochastic neighbor embedding  \citep[t-SNE, ][]{vanDerMaaten2008} is a nonlinear dimensionality reduction technique well-suited for embedding high-dimensional data for visualization in a low-dimensional space of two or three dimensions.} projections of features, offers means of better understanding. If the projection of features from the first fully connected layer (FC 10) of the autoencoder part shows a better separation of type Ia and non-Ia supernovae, than from the seventh convolution layer, the extraction of these features for the contrastive branch degrades the performance. This means that it is preferable to consider a feature representation space of light curves of high abstraction level rather than a representation apparently more suited for classification in the autoencoder layers, as it allows a significant reduction of the mismatch between the training and the test databases. \\ The last layers of the contrastive module  (conv 9c and FC 11c) mark a clear separation between type Ia and non-Ia supernovae (bottom panel of Fig. \ref{tsne_contrastive}).  

\subsubsection{The classification branch}
\begin{figure}
\includegraphics[width=\textwidth / 2]{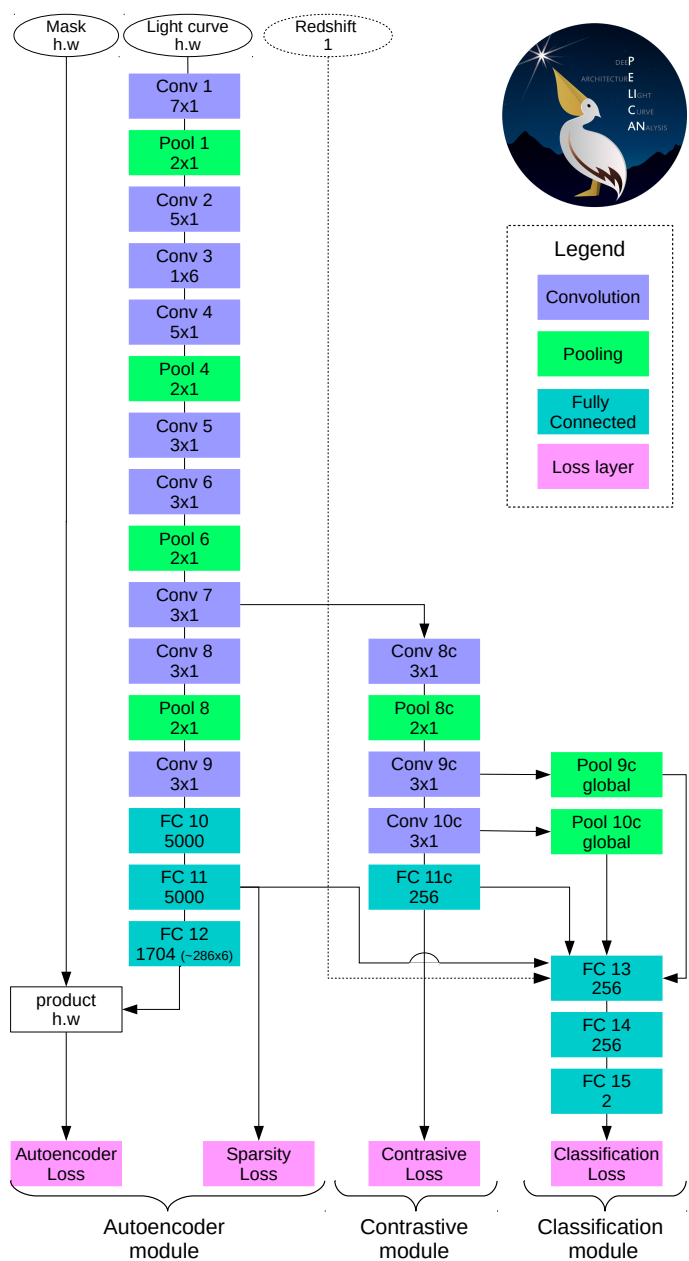}
\caption{Representation of PELICAN architecture which is composed of three modules: the autoencoder, the contrastive and the classification modules. The first module optimizes the autoencoder loss containing a sparsity parameter (see Equation \ref{eq_sparse}). In the second module, the contrastive loss (see Equation \ref{contrast_eq}) is optimized to bring the features with the same label together. Finally the third module performs the classification step optimizing a standard classification loss. }
\label{netork}
\end{figure}

The last module of PELICAN is composed of three fully connected layers (see Fig. \ref{netork}). It takes as input features from the two first modules to perform the classification step. 
Indeed to make the final classification, this part needs information of the first module that fully characterizes light curves of the test database and so give a large variety of features that allows to reduce the mismatch between the training and test databases. However, this time we extract features from the decoder part as it was shown that it is able to make a separation of the classes that is relevant for this final step (see Fig. \ref{tsne_contrastive}). Then the classification branch must benefit from features of the second contrastive branch, and particularly the fully connected layer (FC11c) that reduce again the mismatch while marking a separation between classes. \\ Finally to fight against the overfitting of missing data, the third module takes also as input, features from the ninth and tenth convolution layers of the contrastive branch (conv 9c and conv 10c). We apply a specific operation, called a global pooling, which allows to transform a 2D output feature vector of a convolution layer into a 1D feature vector given as input to a fully connected layer. We choose to apply a global max pooling that will select only the maximum value on the 2D output feature maps from the convolution layers, excluding zero values and so missing data.  \\ We also make use of dropout technique  \citep{dropout} on the two fully connected layers FC13 and FC14 to fight against overfitting.

\section{Light curve data }
We test and adapt our method on three different databases. First we evaluate the techniques on simulated data from the Supernovae Photometric Classification Challenge  \citep[SPCC, ][]{Kessler2010b, Kessler2010a} then on simulated LSST light curves for the main survey and the deep fields. 
Finally we explore the possibility to make the learning step on simulated light curves and then to test on real data. We apply this last work on SDSS supernovae light curves \citep{Frieman2008,Sako2008}.

\subsection{SNANA simulator}
Light curves have been simulated using the SNANA simulator \citep{SNANA}. It is an analysis package for supernovae light curves that contains a simulation, a light curve fitter and a cosmology fitter. It takes into account actual survey conditions and so generate realistic light curves by using the measured observing conditions at each survey epoch and sky location. First the supernovae properties are generated by choosing a shape-luminosity and color parameters, that are used in addition to other internal model parameters to determine the rest-frame magnitude at each epoch. Then K-corrections are applied to transform rest-frame model magnitudes to observed magnitudes in the telescope system. Finally the ideal above atmosphere magnitudes are translated into observed fluxes and uncertainties. Type Ia supernovae light curves are simulated from SALT2 \citep{salt2007A&A...466...11G} or MLCS models \citep[]{Jha2007ApJ...659..122J,SNANA}. But there are no such models for non-Ia types. So the simulations uses a library of spectral templates that give the supernovae flux as a function of epoch and wavelength. Only well-sampled photometric light curves are used because spectral templates are interpolated to cover all wavelengths and epochs. The current library contains composite and individual templates for types Ib, Ibc, IIn and IIP. 

\subsection{SPCC data}
The SPCC dataset is composed of simulated light curves of supernovae in \textit{griz} filters of the Dark Energy Survey (DES). The dataset were subdivided in a spectroscopically confirmed subset of 1,103 light curves, which should constitute the training dataset, and a test dataset of 20,216 light curves. However the training dataset is small and highly biased as it is not representative in brightness and in redshift compared to the test set.

\subsection{Simulated LSST data}
\label{lsst_simu_sec}
As LSST will observe a large amount of supernovae, the photometric classification of supernovae types from multi-band light curves is necessary. There will be two main kind of cadences. The first one dedicated to the main survey is called the Wide-Fast-Deep (WFD). It will scan a very large area of the sky. The second one, called Deep Drilling Fields (DDF), will focus on small part of the sky with a higher cadence and deeper images. Thus this will correspond to well measured light curves (see Fig. \ref{lc}) but for a smaller sample.

To validate our method in the context of the future LSST data we simulated light curves of supernovae as observed with the WFD and DDF observational strategies, with the minion 1016 baseline model \citep{Minionbiswas_rahul_2017_1006719}. The simulation was realized in the \textit{ugriz} filters of the LSST. We assume a $\Lambda$CDM cosmology with $\Omega_M = 0.3156$, $\Omega_\Lambda = 0.6844$ and $w_0=-1$. Simulations are made in a finite redshift range, $z \in [0.05,1.20]$. We consider an efficiency for the image subtraction pipelines reaching 50\% around Signal-to-Noise Ratio (SNR) $\sim 5$. Each object must have 2 epochs in any band. For the simulation of type Ia light curves the color and the light curve shapes parameters vary in the following intervals: $c\in [-0.3,0.5]$, $x_1 \in [-3,2]$. The simulation of non-Ia types is based on a library of spectral templates for types Ib, Ibc, IIn and IIP.\\ Our simulation includes a spectroscopically confirmed sample from the DDF survey. It is based on observations from a 8 m class telescope with a limiting i-band magnitude of 23.5. In this work we assume different allocating time for the spectroscopic follow-up. A reasonable scenario allows a spectroscopic follow-up of 10\% of observed light curves in DDF, i.e 2k spectroscopically confirmed light curves of supernovae. But we also consider a most restrictive case by assuming that only 500 then 1,000 light curves are spectroscopically confirmed. Moreover we explore two ideal cases for which 5k then 10k supernovae have been followed up. Finally we also consider different number of photometric observations of light curves as it is interesting to classify light curves before 10-years observation of LSST.  All the configuration are summarized on Table \ref{tablo_final}.\\

\subsection{SDSS data}
As simulated data do not reproduce perfectly the real data, it is interesting to test our method on real data. The ideal strategy is to simulate light curves that corresponds to SDDS survey to train the model and then test on real data. This is a challenging methodology as there is a severe mismatch between the training and the test databases. However making a model able to remove this kind of mismatch is crucial for the future surveys where the spectroscopic follow-up is limited. 
Therefore we simulated light curves of supernovae that corresponds to SDSS data.
Then, we extracted light curves in \textit{ugriz} filters from the SDSS-II Supernova Survey Data \citep{Frieman2008,Sako2008}. The SDSS-II SN data were obtained during three month campaigns in the Fall of 2005, 2006 and 2007 as part of the extension of the original SDSS. The Stripe 82 region was observed with a rolling cadence. Some spectroscopic measurements were performed for promising candidates depending on the availability and capabilities of telescopes \citep{Sako2008}. A total of 500 SN Ia and 82 core collapse SN were spectroscopically confirmed.

\section{Experimental protocol}

\label{exp}
In this section, we explained the protocol and the different techniques used for the training process.

\subsection{Data augmentation}
\label{data_aug}

In this classification context, data augmentation is a crucial step. Indeed, in order to make PELICAN robust against the differences between the training and the test databases (i.e sampling, mean magnitude, noise...), it is essential to use different data augmentation techniques. Moreover when light curves that composed the training and test databases are measured with different observational strategies the difference in sampling is increased  and the data augmentation has to be reinforced. It is the case in the context of LSST if we compare light curves from the WFD survey on the one hand, and light curves from the DDF survey on the other hand. To make PELICAN able to learn on DDF light curves and generalize on WFD light curves, the data augmentation has to be adapted. \\ Finally as supernovae from the test database are often fainter, errors on their fluxes are often bigger. Therefore the data augmentation needs also to be applied to the errors. \\ Thus, in addition to the uniform noise applied differently on each magnitude of light curves given as input to the denoising autoencoder, we add two other kind of noise on magnitudes of light curves:  

\begin{itemize}
\item an uniform constant noise $\in [-1.0,1.0]\, \textrm{mag}$ which is added to all the magnitudes of the light curve,
\item an uniform noise $\in [0.93,1.07]$ which is multiplied by all the magnitudes of the light curve,
\end{itemize}

Then we randomly remove one or several magnitudes or/and all magnitudes for a given band. This process is particularly effective for the classification of light curves of supernovae observed with a WFD strategy based on a training on supernovae light curves from the DDF survey.\\
Finally, to prevent PELICAN model from learning the missing value positions on each light curve, we perform random time-translations keeping all the data points but varying their positions in time. So the learning becomes invariant to the position of points.

\subsection{Setting learning parameters}
We used the Adam optimizer \citep{Adam} for all the training steps in different modules with a learning rate decreasing by a 10 factor after 25 000 iterations. The batch size during the learning is fixed to 96 light curves. \\
For the autoencoder learning, we optimized the values of the sparsity parameters over one validation base and use them for all the different configurations, as it is not sensitive to the database. We set $\rho_a$ and $\rho_b$ equal to $5 \times 10^{-4}$ and $0.0$ respectively and $\alpha$ to $0.01$. \\
The cut parameter $m^{c}$ in the i-band magnitude from the second module, depends on the database. We choose its value in order to have enough examples on both sides of the cut in magnitude. We set $m^{c}$ to 23.5 mag, 24.5 mag and 22.5 mag for the SPCC, LSST and SDSS databases respectively. The values of these parameters are not sensitive and a small variation of them did not change the results.

\subsection{Ensemble of classifiers}
To increase the performance, we trained an ensemble of classifiers as it was shown to be more accurate than individual classifiers \citep[e.g.][]{ensemble}. Moreover the generalization ability of an ensemble is usually stronger than that of base learners. This step involves training N times one model with the same training database but a different initialization of the weights. We chose N=7 and the individual decisions were then averaged out to obtain the final values of probabilities. This step allows to increase the accuracy of $2\%$ on average.

\section{Results}
In this section we present the results that we obtained for each dataset. 

\subsection{Metrics}
\label{metric}
To evaluate the performance of PELICAN in different contexts we use several commonly used statistic metrics that are the Accuracy (Acc), the recall (R) or true positive rate (TPR), the precision (P) and the false positive rate (FPR). They are defined from the following confusion matrix:

\begin{tabular}{l|l|c|c|c}
\multicolumn{2}{c}{}&\multicolumn{2}{c}{Predictive label}&\\
\cline{3-4}
\multicolumn{2}{c|}{}&Ia&Non Ia\\
\cline{2-4}
\multirow{2}{0.8cm}{True label }& Ia & $\small{\textrm{True Positive (FP)}}$ & $\small{\textrm{True Negative (TN)}}$ \\
\cline{2-4}
& Non Ia & $\small{\textrm{False Positive (FP)}}$ & $\small{\textrm{True Negative (TN)}}$ \\
\cline{2-4}
\end{tabular}

\begin{align}
\hspace{30pt}\llap{\textbullet\hspace{10pt}} &Acc = \frac{TP+TN}{(TP+FP+TN+FN) }  \\
\hspace{30pt}\llap{\textbullet\hspace{10pt}} &R \,\, (\textrm{or } TPR) = \frac{TP}{(TP+FN)} \\
\hspace{30pt}\llap{\textbullet\hspace{10pt}} &P  = \frac{TP}{(TP+FP)} \\
\hspace{30pt}\llap{\textbullet\hspace{10pt}} &FPR  = \frac{FP}{(FP+TN)} 
\end{align}
\\
As a graphical performance measurement we use the ROC (Receiver Operating Characteristic) curve which is plotted with TPR on y-axis against FPR on x-axis. It gives an estimation of the performance of a classifier at different thresholds settings. The best possible prediction method would yield a point in the upper left corner or coordinate (0,1) of the ROC space, representing the lack of false negatives and false positives. A random guess would give a point along a diagonal line from the left bottom to the top right corners.
\\
From the ROC graphic, we can extract the value of the AUC (Area Under the Curve) which captures the extent to which the curve is up in the upper left corner. The score has to be higher than 0.5 which is no better than random guessing.

\subsection{SPCC}
The first evaluation of PELICAN is made on the SPCC dataset. We trained the model with two different training datasets: a representative training database and a non-representative training dataset. The representative training database is a simplified theoretical scenario, in which there is no limitation in brightness and redshift of the spectroscopic follow-up. It is built by selecting randomly 1,103 light curves from the whole dataset. The non-representative training database, which represents the real scenario, is the spectroscopically confirmed subset of 1,103 light curves that was proposed for the challenge. This last one is non-representative of the test dataset in brightness and redshift. 

\begin{figure*}[h!]
\center
\includegraphics[width=19cm]{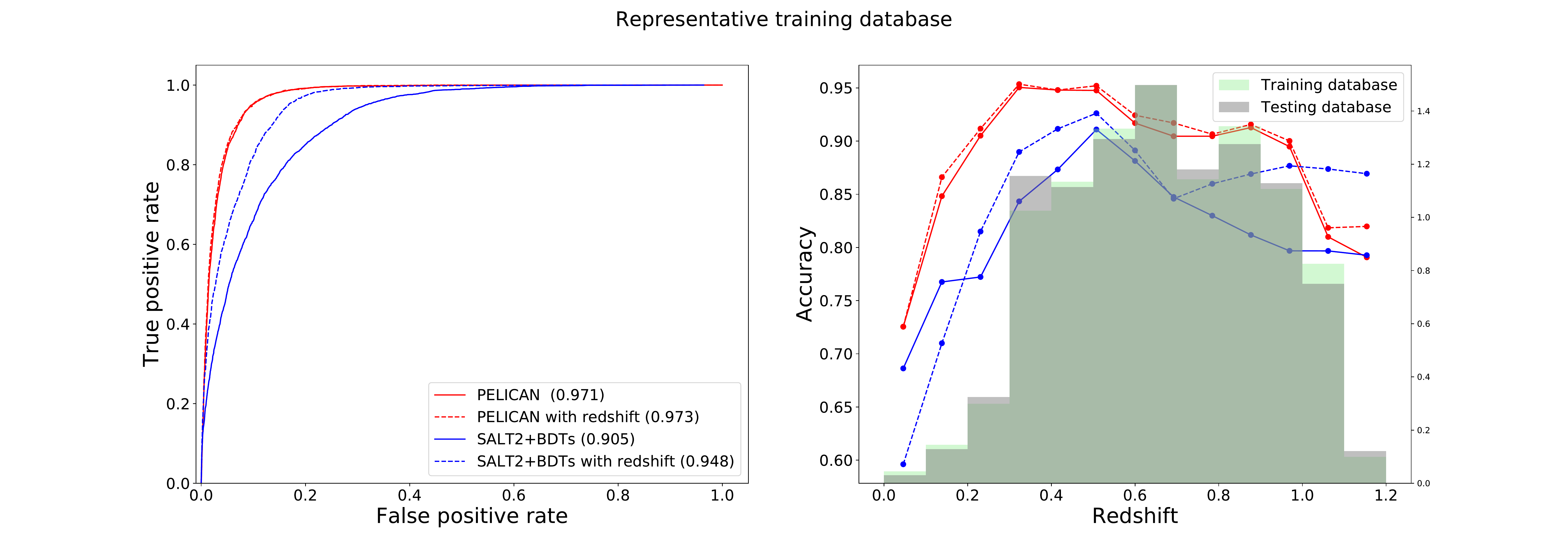}
\includegraphics[width=19cm]{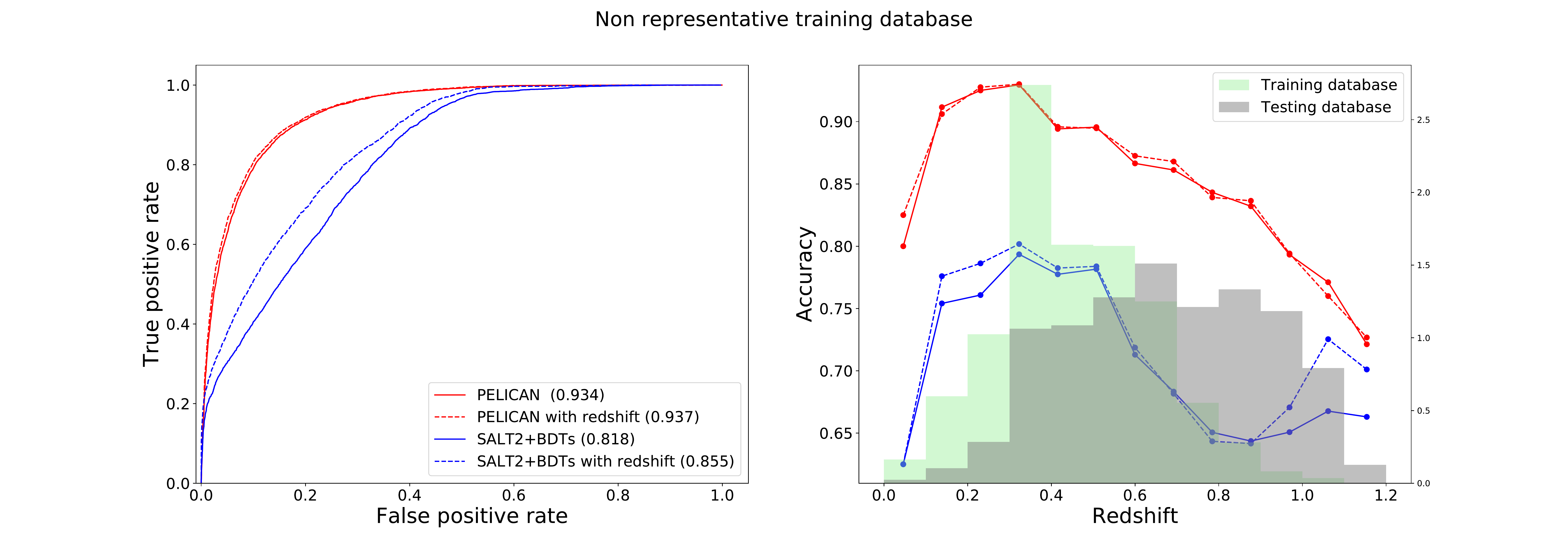}
\caption{Comparison of ROC curves with the AUC score in brackets (left panels) and the accuracy versus redshift (right panels) for PELICAN (in red) and BDTs method (in blue), with (solid lines) and without (dashed lines) the redhift inside the training. The representative case is on the first line and the non-representative one on the second line.}
\label{perf_challenge}
\end{figure*}

As shown in Lochner et al. (\citeyear{Lochner2016}, noted L16 hereafter) the best average AUC is obtained by extracting SALT2 features and using boosted decision trees (noted BDTs hereafter) as classifier. Therefore we compared the performance of PELICAN with BDTs algorithm that take as input SALT2 features. We test both methods  with and without the information of the redshift inside the training. \\
The ROC curves for both methods are represented on Fig. \ref{perf_challenge} (on left panels) and values of statistics are reported in Table \ref{challenge}.  \\
By considering a non-representative training database, without including the redshift during the training, PELICAN obtains an accuracy of 0.856 and an AUC of 0.934 which outperforms BDTs method which reaches 0.705 and 0.818.
If we train both methods on a representative training database, as expected, the performance increases. The accuracy and the AUC become 0.911 and 0.970 with PELICAN, against 0.843 and 0.905 with BDTs algorithm. It is interesting to note that the gain in statistics obtained with PELICAN, is lower than of BDTs values, which means that PELICAN is able to better deal with the problem of mismatch. This ability will be confirmed by the promising results obtained with a non-representative training database composed of LSST light curves simulations (see Section \ref{lsst_result}).\\ The performance of PELICAN does not change by adding the redshift information during the training, which is not the case for BDTs algorithm, for which the accuracy and the AUC are slightly increased. This might means that PELICAN is able to extract by itself the redshift information during its training. Figure \ref{perf_challenge} shows the accuracy as a function of redshift, with the corresponding redshift distributions and BDTs results for comparison. If PELICAN is trained on a representative training database, the accuracy tends to decrease at low redshifts and at redshift above 1.0, as the corresponding redshift bins are poorly populated at these extreme values. A further trend is observed for BDTs, except at redshift above 1.0, and only if redshift values are included as an additional feature for the training. By considering a non-representative training database, the accuracy significantly decreases at high redshift for both methods. As the addition of redshift inside the training does not change the tendency obtained by PELICAN, this trend in function of redshift is likely due to the too small number of examples at high redshifts.

\begin{table}
\begin{tabular}{|>{\centering}m{2.25cm}|>{\centering}m{1.1cm}|>{\centering}m{1.4cm}|c|>{\centering}m{1.1cm}|>{\centering}m{1.1cm}|}
\hline 
 \textbf{SPCC} & {\small{}Redshift} & {\small{}Training database}  & {\small{}Accuracy} & {\small{}AUC}\tabularnewline
\hline 
\multirow{2}{2.25cm}{{\footnotesize{}\specialcell{Non-representative \\ training database \\ \tiny{(SALT2+BDTs)}}} } & {\small{}no} & 1103 {\tiny{ \,\,spec}}  & {\small{}  \specialcell{0.856\\\tiny{(0.705)}} } & {\small{} \specialcell{0.934 \\\tiny{(0.818)}} }\tabularnewline
\cline{2-5} 
 & {\small{}yes} & 1103 {\tiny{\,\,\,\,spec}} & {\small{} \specialcell{0.863\\\tiny{(0.713)}} }& {\small{} \specialcell{0.939\\\tiny{(0.855)}}}\tabularnewline
\hline 
\hline 
\multirow{2}{2.25cm}{{\footnotesize{}\specialcell{Representative \\ training database \\ \tiny{(SALT2+BDTs)}}}} & {\small{}no} & 1103 \\ {\tiny{mix of spec and phot}}  & {\small{} \specialcell{0.911 \\ \tiny{(0.843)}}} & {\small{} \specialcell{0.970 \\ \tiny{(0.905)}}}\tabularnewline
\cline{2-5} 
 & {\small{}yes} & 1103 \\ {\tiny{mix of spec and phot}}  & {\small{} \specialcell{0.917 \\ \tiny{(0.878)}} }& {\small{} \specialcell{0.971 \\ \tiny{(0.948)}}}\tabularnewline
\hline 
\end{tabular}
\caption{Statistics obtained for SPCC challenge by PELICAN and BDTs results in parenthesis. The first part reports results for a non-representative training database and the second part for a representative training database. We consider both cases by adding or not the redshift values inside the training.}
\label{challenge}
\end{table}

\subsection{LSST simulated light curves}
\label{lsst_result}
The next step is to evaluate PELICAN on simulated LSST light curves under realistic conditions. In this way, we consider for all the tests a non-representative spectrocopically confirmed database from the DDF survey as represented on Fig. \ref{mismatch}. We consider different configurations of the training and test databases. We constrain the number of spectroscopically confirmed light curves to vary between 500 light curves to 10k. Even if the upper bound corresponds to an ideal scenario in which roughly more than 40\% of light curves in the DDF have been following up, it is interesting to compare the performance of PELICAN with large training sample.

\begin{figure}[h!]
\center
\includegraphics[width=9cm]{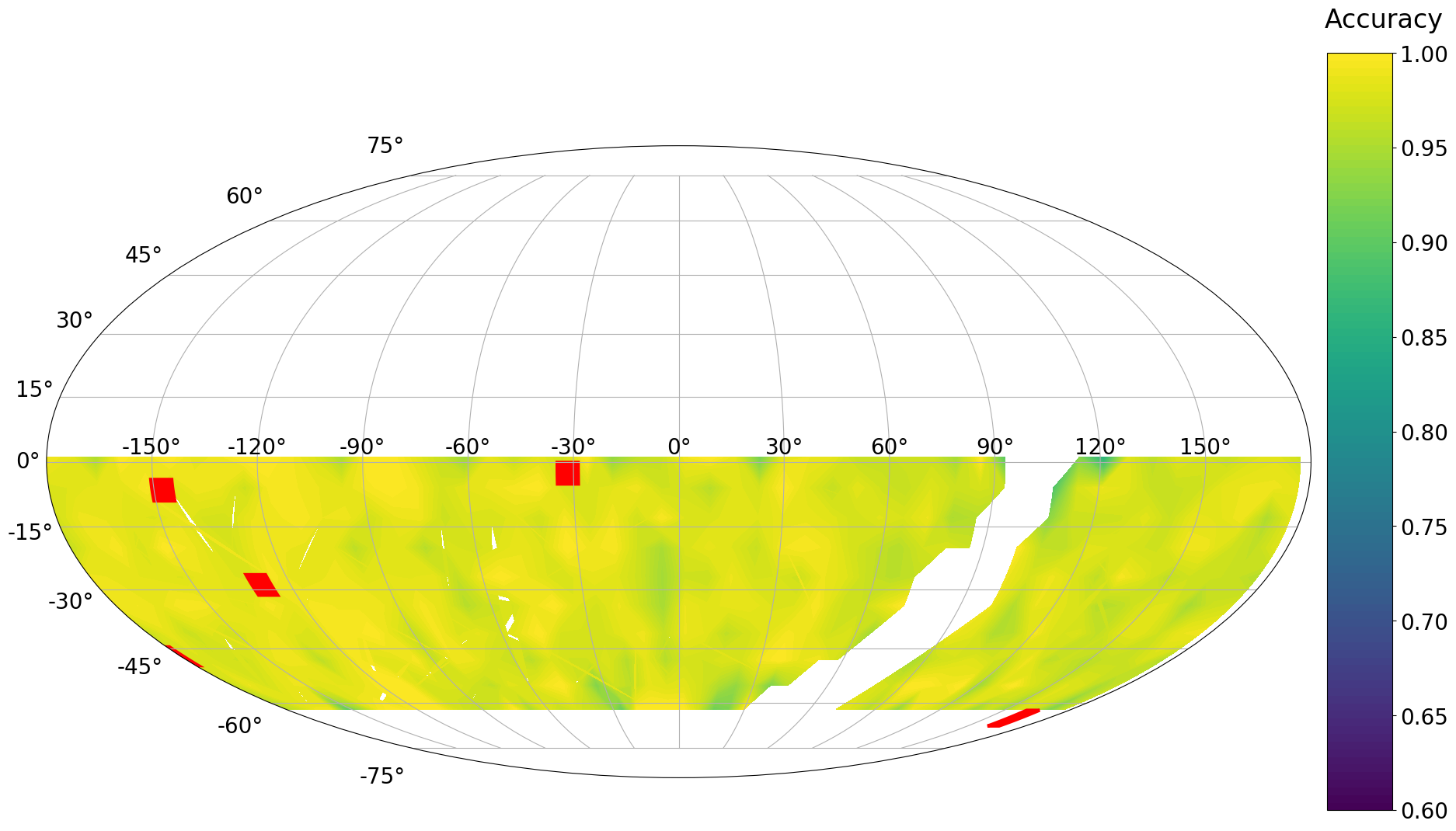}
\caption{Spatial distribution of the accuracy obtained with PELICAN for the classification of light curves simulated with minion 1016 cadence model. The Deep Drilling Fields are represented by red squares.}
\label{map_kraken_2026}
\end{figure}

We simulated light curves with the minion 1016 cadence model. This model includes a WFD survey and five DDF (see Fig. \ref{map_kraken_2026}) {\color[red]{A VERIFIER!} } 
It is not certain that a spectroscopic follow-up will be performed on supernovae light curves in WFD fields. So we use a different approach which consists to train PELICAN on DDF light curves and then adapt the pre-trained model to classify supernovae light curves observed in WFD survey. This strategy allows to consider the possibility to benefit from SN Ia candidates from WFD fields to constrain the cosmological parameters, without any spectroscopic follow-up of the main survey. 

\subsubsection{Classification of DDF light curves}
\begin{figure*}[h!]
\center
\includegraphics[width=19cm]{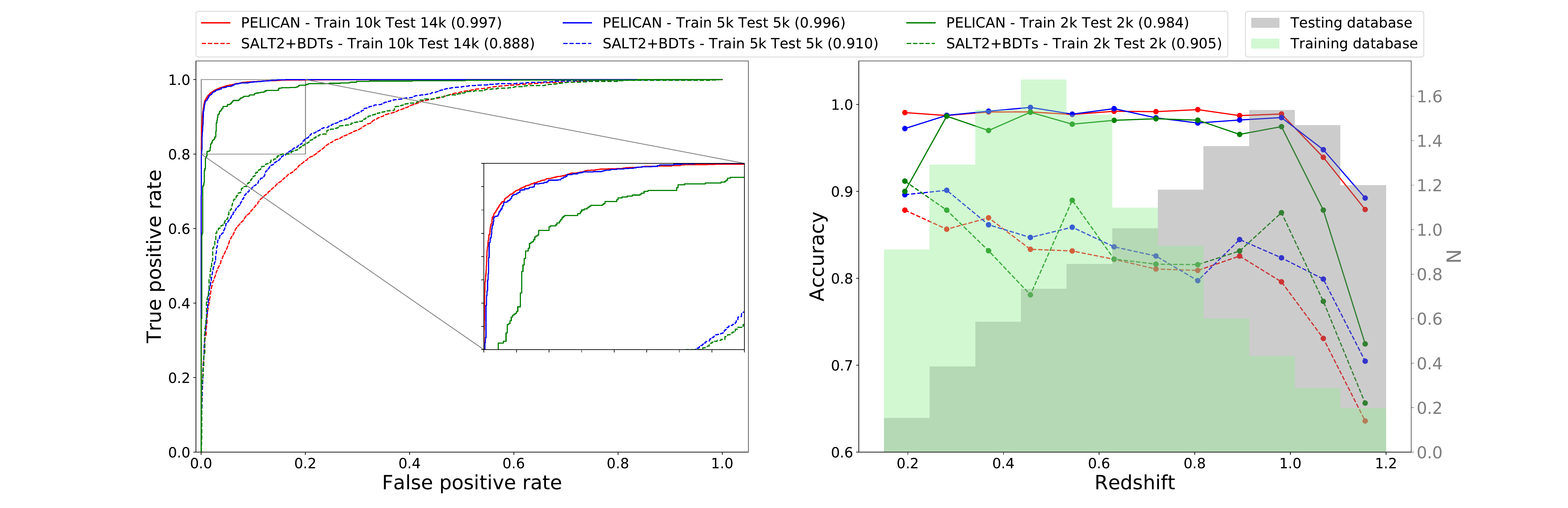}
\caption{Comparison of ROC curves for different training configurations of DDF survey, with the AUC score in brackets (left panel) and the accuracy versus redshift (right panel) for PELICAN (in solid lines) and BDTs method (in dashed lines). }
\label{lsst_red}
\end{figure*}

The results of the different configurations are reported in Table \ref{tablo_final} and the ROC curves for some of these configurations on Fig. \ref{lsst_red}. In addition to the values of the accuracy and the AUC, we compare the recall of SN Ia by constraining the precision to be higher than $95\%$ and $98\%$. Such a level of contamination becomes competitive with spectroscopy contamination. Again we compared the performance of PELICAN with the best method highlighted in L16, that is BDTs and SALT2 features. Even if this method  was not designed for such training configurations, it allows to compare a feature-based machine learning method to PELICAN.  \\ If we consider the most constraining configuration composed of only 500 spectrocopically confirmed light curves for the training and 1,500 light curves for the test database, PELICAN reaches an accuracy of 0.895, and an AUC of 0.966.  Moreover PELICAN is then able to detect 76.9\% of SN Ia with a precision higher than 95\% and 60.2\% with a precision higher than 98\%. These results are quickly improved by considering more examples on both training and test databases. The number of light curves inside the test database is important, especially if the number of examples in the training database is small, as the autoencoder is trained on the test database. Indeed there is about a 8\% improvement factor of the recall by going from 1k to 3k light curves in the test database with a fixed number of examples in the training database of 1k light curves. But this factor becomes negligible if the number of spectroscopic confirmed light curves is sufficient, i.e from 5k examples, with an improvement of around $0.2\%$. \\
The configuration that seems reasonable after 10 years of observation includes a spectroscopic follow-up of 10\% of the observed light curves, i.e 2k light curves of supernovae, and a test database of 22K light curves. For this realistic scenario, PELICAN reaches an accuracy of 0.942 and is able to correctly classify 87.4\% of SN Ia with a precision higher than 98\%, which constitutes a major result of our study meaning that a large fraction of SN Ia are well-classified by PELICAN, with a precision comparable to a spectroscopy measurement. By considering 10k light curves in the training database, the number of detected SN Ia is then increased by 9\%. All results obtained by PELICAN outperform those obtained by BDTs (the BDTs values are listed in parenthesis in Table \ref{tablo_final}).\\

The right panel of Fig. \ref{lsst_red} shows the accuracy as a function of redshift, with the corresponding redshift distributions on both training and test databases, and the BDTs results for comparison. The accuracy of PELICAN does not depend on redshift until 1.0 where it slightly decreases. This tendency is likely due to the small number of training examples at redshift higher than 1.0. BDTs method shows the same behaviour at high redshifts.

\begin{table*}

\begin{tabular}{|>{\raggedright}m{0.3cm}|>{\centering}m{2.0cm}|>{\centering}m{3.1cm}|>{\centering}m{2.6cm}|>{\centering}m{2cm}|c|c|>{\centering}m{1.25cm}|}
\cline{2-8} 
\multicolumn{1}{>{\raggedright}m{0.3cm}|}{} & Total (LC) & \specialcell{Training database \\ (Spec only)} & \specialcell{test database \\ (Phot only)} & Accuracy & \specialcell{$\textrm{Recall}_{Ia}$ \\\tiny{$\textrm{Precision}_{Ia} > 0.95$}}  & $\specialcell{$\textrm{Recall}_{Ia}$ \\\tiny{$\textrm{Precision}_{Ia} > 0.98$}} $ & AUC\tabularnewline
\hline 
\multirow{11}{0.3cm}{\begin{turn}{90}\centering{LSST DDF \qquad\qquad\qquad\qquad\qquad}\end{turn}} & \multirow{2}{2.4cm}{\centering{} \\2,000 } & $500$ & $1,500$ & $\specialcell{0.895\\ \tiny{(0.795)}}$ & $\specialcell{0.769\\ \tiny{(0.382)}}$  & $\specialcell{0.602\\ \tiny{(0.183)}}$ & $\specialcell{0.966\\ \tiny{(0.885)}}$ \tabularnewline
\cline{3-8} 
 &  & $1,000$ & $1,000$ & $\specialcell{0.917\\ \tiny{(0.888)}}$ &  $\specialcell{0.827\\ \tiny{(0.505)}}$ & $\specialcell{0.687\\ \tiny{(0.353)}}$ & \specialcell{0.973\\ \tiny{(0.898)}} \tabularnewline
\cline{2-8} 
 & \multirow{2}{2.4cm}{\centering{} \\ 4,000 } & $1,000$ & $3,000$ & $\specialcell{0.928\\ \tiny{(0.850)}}$ &  $\specialcell{0.890\\ \tiny{(0.490)}}$ &$\specialcell{0.764\\ \tiny{(0.246)}}$& $\specialcell{0.979\\ \tiny{(0.895)}}$\tabularnewline
\cline{3-8} 
 &  & $2,000$ & $2,000$ & $\specialcell{0.938\\ \tiny{(0.813)}}$ & $\specialcell{0.927\\ \tiny{(0.594)}}$ & $\specialcell{0.806\\ \tiny{(0.256)}}$ & $\specialcell{0.984\\ \tiny{(0.905)}}$\tabularnewline
\cline{2-8} 
 & \multirow{3}{2.4cm}{\centering{} \\ 10,000 } & $2,000$ & $8,000$ & $\specialcell{0.946\\ \tiny{(0.796)}}$ & $\specialcell{0.944\\ \tiny{(0.496)}}$ & $\specialcell{0.886\\ \tiny{(0.284)}}$ & $\specialcell{0.989\\ \tiny{(0.891)}}$\tabularnewline
\cline{3-8} 
 &  & $3,000$ & $7,000$ & $\specialcell{0.950\\ \tiny{(0.809)}}$ & $\specialcell{0.950\\ \tiny{(0.548)}}$ &  $\specialcell{0.903\\ \tiny{(0.285)}}$ & $\specialcell{0.990\\ \tiny{(0.905)}}$\tabularnewline
\cline{3-8} 
 &  & $5,000$ & $5,000$ & $\specialcell{0.971\\ \tiny{(0.818)}}$ & $\specialcell{0.981\\ \tiny{(0.510)}}$ & $\specialcell{0.959\\ \tiny{(0.315)}}$ & $\specialcell{0.996\\ \tiny{(0.910)}}$\tabularnewline
\cline{2-8} 
 & \multirow{4}{2.4cm}{\centering{}\\ 24,000 } & $2,000$ & $22,000$ & $\specialcell{0.942\\ \tiny{(0.792)}}$ & $\specialcell{0.940\\ \tiny{(0.477)}}$ & $\specialcell{0.874\\ \tiny{(0.209)}}$ & $\specialcell{0.986\\ \tiny{(0.890)}}$ \tabularnewline
\cline{3-8} 
 &  & $3,000$ & $21,000$ & $\specialcell{0.945\\ \tiny{(0.797)}}$ &  $\specialcell{0.937\\ \tiny{(0.474)}}$ & $\specialcell{0.891\\ \tiny{(0.254)}}$ & $\specialcell{0.986\\ \tiny{(0.892)}}$\tabularnewline
\cline{3-8} 
 &  & $5,000$ & $19,000$ & $\specialcell{0.968\\ \tiny{(0.805)}}$ &  $\specialcell{0.978\\ \tiny{(0.485)}}$ & $\specialcell{0.957\\ \tiny{(0.228)}}$ & $\specialcell{0.996\\ \tiny{(0.898)}}$ \tabularnewline
\cline{3-8} 
 &  & $10,000$ & $14,000$ & $\specialcell{0.971\\ \tiny{(0.790)}}$ & $\specialcell{0.983\\ \tiny{(0.465)}}$ & $\specialcell{0.965\\ \tiny{(0.260)}}$  & $\specialcell{0.997\\ \tiny{(0.888)}}$ \tabularnewline
\hline 
\hline 
\multirow{3}{0.4cm}{\begin{turn}{90}
\centering{}LSST WFD\,\, \end{turn}}
\rule{0pt}{15pt} & 17,000 & \footnotesize{DDF Spec : $2,000$} & \footnotesize{\specialcell{ WFD : $15,000$}} &\specialcell{0.965\\ \tiny{(0.620)}} & \specialcell{0.982\\ \tiny{(0.041)}} &  $\specialcell{0.905\\ \tiny{(0.008)}}$ & \specialcell{0.992\\ \tiny{(0.703)}} \tabularnewline
\cline{2-8} \rule{0pt}{15pt} 
 & 43,000 & \footnotesize{DDF Spec : $3,000$} & \footnotesize{\specialcell{WFD : $40,000$}} & \specialcell{0.976\\ \tiny{(0.623)}} & \specialcell{0.995\\ \tiny{(0.018)}} &  $\specialcell{0.964\\ \tiny{(0.000)}}$ & \specialcell{0.996\\ \tiny{(0.711)}}\tabularnewline
\cline{2-8} \rule{0pt}{15pt} 
 & 90,000 & \footnotesize{DDF Spec : $10,000$ } & \footnotesize{\specialcell{WFD : $80,000$}} & \specialcell{0.978\\ \tiny{(0.620)}} & \specialcell{0.995\\ \tiny{(0.046)}} & $\specialcell{0.973\\ \tiny{(0.000)}}$ & \specialcell{0.997\\ \tiny{(0.709)}} \tabularnewline
\hline 
\end{tabular}
\caption{Statistics for various training configurations on the DDF survey (first part) and the WFD survey (second part), with BDTs results in parenthesis. The different metrics are defined in Section \ref{metric}. }
\label{tablo_final}
\end{table*}

\begin{table}
\begin{tabular}{|>{\centering}m{2.8cm}|>{\centering}m{2.5cm}|c|>{\centering}m{0.75cm}|>{\centering}m{0.8cm}|}
\hline 
{\small{}Training database} & {\small{}test database}  & {\small{}Accuracy} & {\small{}AUC}\tabularnewline
\hline 
{{\footnotesize{}{}SDSS simulations : 219,362}} & {\footnotesize{}SDSS-II SN confirmed : 582}  & {\small{}  \specialcell{0.462} } & {\small{} \specialcell{0.722}} \tabularnewline
\hline 
{\footnotesize{}\specialcell{SDSS simulations : \\ 219,362  \\ SDSS-II SN confirmed : \\ 80}} & {\footnotesize{}SDSS-II SN confirmed : 582}  & {\small{}  \specialcell{0.868} } & {\small{} \specialcell{0.850}} \tabularnewline
\hline 
\end{tabular}
\caption{Statistics obtained on real SDSS data. The first line reports results obtained by training PELICAN only on simulated data. By including only 80 SDSS light curves in the training results are significantly improved (second line).  }
\label{sdss_table}
\end{table}

\subsubsection{Classification of light curves in WFD}
\begin{figure*}[h!]
\center
\includegraphics[width=19cm]{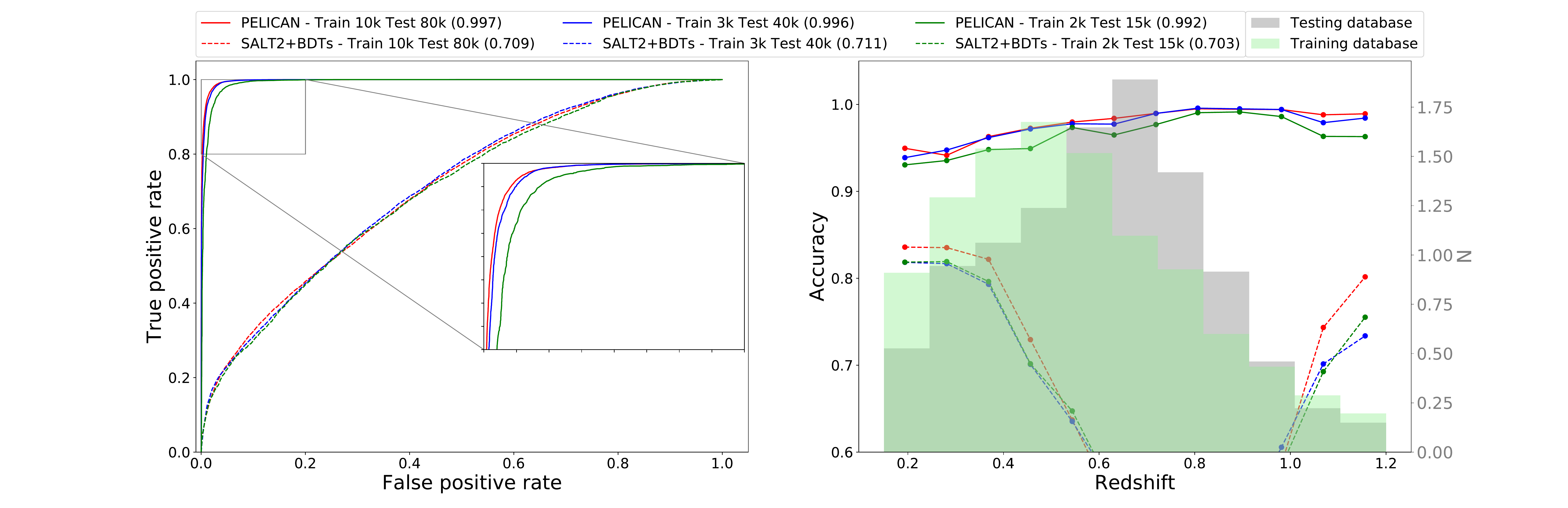}
\caption{Comparison of ROC curves for different training configurations of WFD survey, with the AUC score in brackets (left panel) and the accuracy versus redshift (right panel) for PELICAN (in solid lines) and BDTs method (in dashed lines).}
\label{wfd_red}
\end{figure*}

The spectroscopic follow-up of SNe Ia candidates is uncertain on WFD survey. Nevertheless to increase statistics of SNe Ia for cosmological studies, it is interesting to make PELICAN able to classify supernovae light curves from WFD survey. The strategy consists to train PELICAN on DDF light curves and then test on light curves observed on WFD fields. However this methodology leads to another kind of mismatch over and above the existing mismatch between spectroscopically confirmed light curves and unconfirmed ones. Indeed the unconfirmed supernovae from the DDF survey have a different cadence and observational conditions than those of WFD survey. So the present mismatch is largely increased between the training and test databases. The non-supervised step allows to reduce it as it does for the classification of DDF light curves, but it is not sufficient. The other needed ingredient is the data augmentation to make DDF light curves looking like WFD light curves. Thus we performed a severe data augmentation as WFD light curves are about an average of four times more sparse in \textit{u} and \textit{g} bands, and 1.5 times in \textit{r}, \textit{i} and \textit{z} bands. So we randomly removed until 85\% of observations of each DDF light curve depending on each band.\\
Results are reported on Table \ref{tablo_final} and ROC curves on Fig. \ref{wfd_red}. We consider three configurations of the training and test databases. First we trained PELICAN on a training database of 2k light curves which could constitute a realistic scenario in which 10\% of supernovae in DDF have been spectroscopically confirmed after 10 years of observation. We also consider a training database composed of 3k supernovae light curves from DDF as it is still a realistic scenario which includes a spectroscopic follow-up of 12.5\% of supernovae in DDF survey. Finally we trained PELICAN on an ideal training database of 10k supernovae light curves. \\
With only 2k DDF light curves for the training database and 15k light curves for the test database, PELICAN reaches an accuracy of 0.965. It is able to classify 98.2\% of supernovae with a precision higher than 95\% and 90.5\% with a precision higher than 98\%. If we consider 3k light curves for the training database and 40k for the testing database, the percentage of well classified light curves, with a precision higher than 98\%, is 96.4\%.
With 10k light curves in the training database and 80k in the testing database, the improvement factor is about 1\%, where 97.3\% of supernovae Ia are correctly classified by PELICAN with a precision higher than 98\%. These encouraging results open the perspective to use SN Ia candidates from WFD survey, whose the classification precision is comparable to a spectroscopic identification, to constrain cosmological parameters. \\
BDTs method obtained poor results for this complex training configuration. This kind of feature-based algorithm have to be adapted to overcome the problem of mismatch which significantly degrade the performance.\\

The accuracy as a function of redshift shows a behaviour that might seem strange (see Fig. \ref{wfd_red}). Indeed, the accuracy slightly increases with redshift. This bias is probably due to the mismatch of redshift distributions between the DDF and WFD lights curves. Indeed, during the non-supervised step, low redshifts examples are under-represented in the test database, that causes a decrease of the accuracy. 
This strange behaviour is increased for BDTs results. Indeed if the accuracy decreases until redshift 1.0, it increases at redshift above 1.0. This significant bias is due to the mismatch between the DDF and WFD light curves. Indeed BDTs algorithm was not designed to deal with this kind of training configuration. 

Figure \ref{map_kraken_2026} exhibits no bias across the sky as the accuracy is uniform on WFD survey.

\subsection{Classification of real SDSS light curves}
The last evaluation of PELICAN is made on real data. The stakes are important as developing a method that can be trained on simulated data while reaching good performance on real data, offers great opportunity for the future surveys. Indeed by using simulated data, the size of the training database could be unlimited and the problem of mismatch between the training database and the test database could be removed.
We evaluate PELICAN only on spectroscopically confirmed supernovae that corresponds to 500 SN Ia and 82 core collapse supernovae. In the first step we train PELICAN only on simulated data and test on real SDSS light curves but it reaches poor performance due to the mismatch between simulated and real data (see Table \ref{sdss_table}). Indeed the sampling and the noise are ideal on simulated data but it is not the case for real ones.  
But if we add only 80 real light curves during the training of PELICAN in order to learn features from real data, the performance are significantly improved with an accuracy of 0.868 and an AUC of 0.850. 
This is a promising result as with only a small subsample of real light curves PELICAN can be trained only on simulated data and reaches good performance on real data.

\section{Further analysis of the behaviour of PELICAN }
In this Section, we study the impact of characteristics of the input LSST simulated light curves relating to properties or observing conditions.

\begin{figure*}[h!]
\center
\includegraphics[width=18cm]{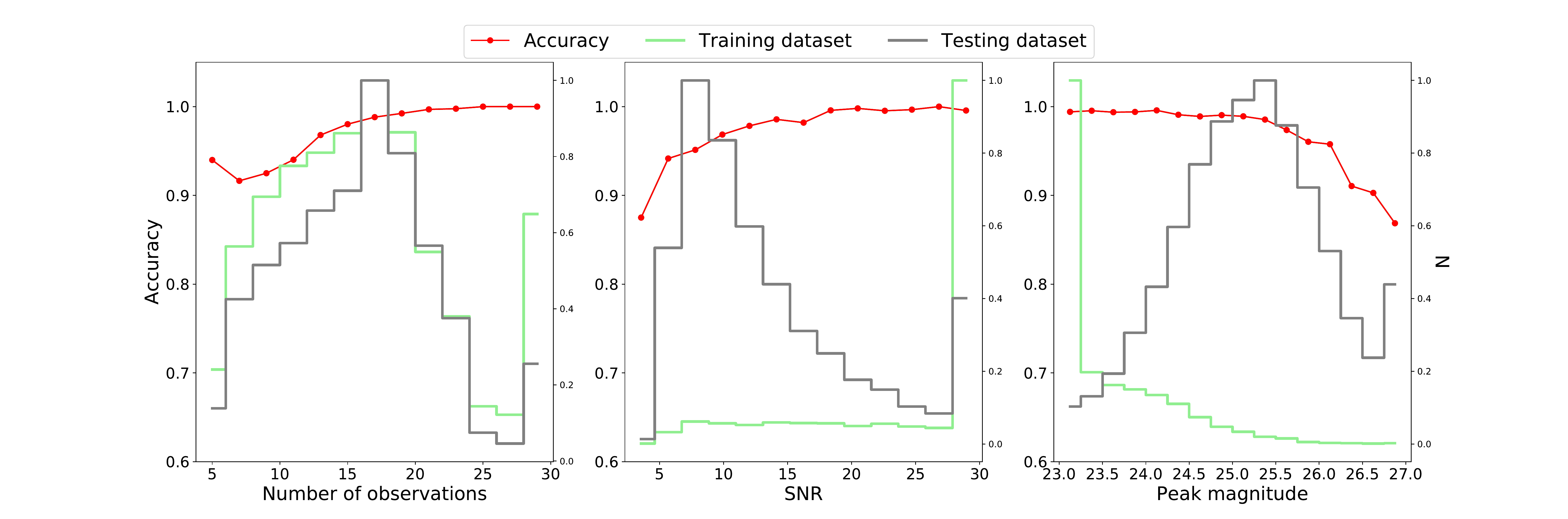}
\includegraphics[width=18cm]{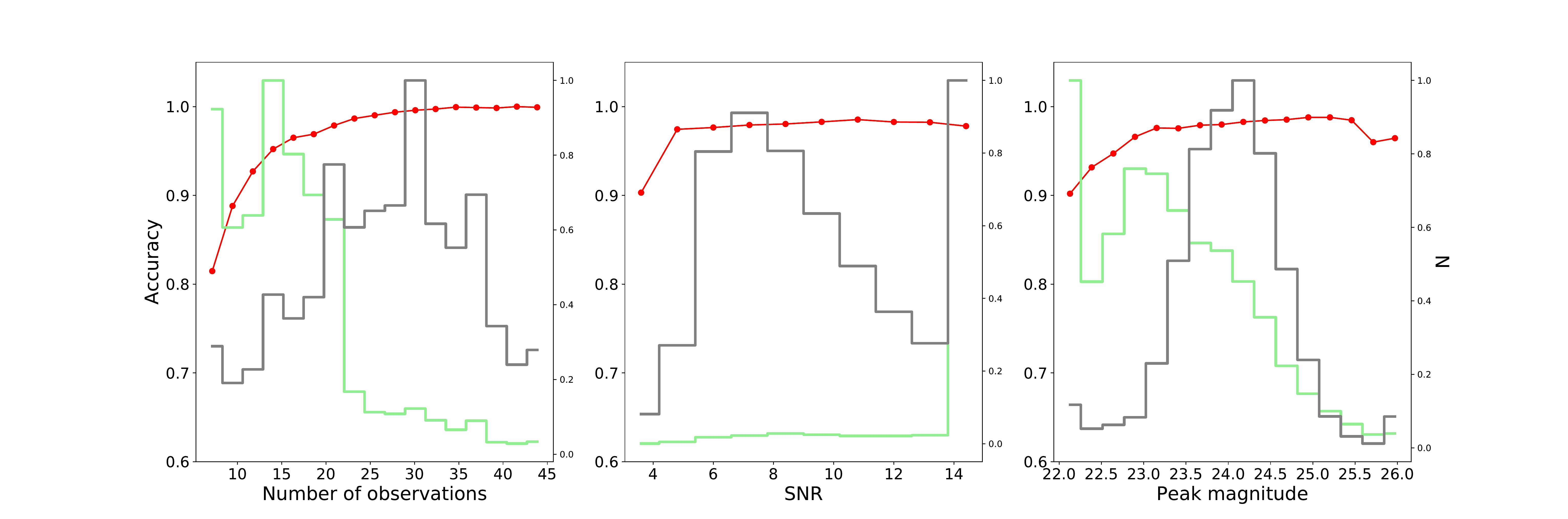}

\caption{The uper panels corresponds to the classification of DDF light curves and the lower panels to the classification of WFD light curves. \textbf{Left panels} : Accuracy as a function of the total number of observations for all bands.\textbf{ Middle panels}: Accuracy as a function of SNR, which is computed as the maximum SNR from all bands. \textbf{Right panels}: Accuracy as a function of peak magnitude, which corresponds to the maximum value of peak magnitude from all bands.  For each case the distribution of the training database is represented in light green and that of the test database in grey.}
\label{prop}
\end{figure*}

\subsection{Influence of the number of observations}
As PELICAN takes as input only light curves, the method should depend on the number of observations. Figure \ref{prop} (left panels) shows the correlation between the number of observations on the light curve in all bands, and the accuracy. For the classification of DDF light curves, the accuracy decreases by a small factor of about 9\%, as the distributions of the number of observations are the same in both the training and test databases. However, for the classification of WFD light curves, the mismatch is present as PELICAN is trained on DDF light curves that have more observations. So this non-representativeness leads to a further decline of the accuracy, of about 20\%.

\subsection{Effect of noise}
We analyze the impact of SNR, computed as the maximum SNR from all bands,  on the accuracy of PELICAN (see middle panels of Fig. \ref{prop}). For the classification of DDF light curves, the accuray decreases at small SNR of roughly 10\%. For the classification of WFD light curves, this trend has reduced and PELCIAN is more robust at low SNR. This is probably due to the first step of non-supervised learning where PELICAN has "seen" light curves with a low SNR, and the data augmentation that we performed. Indeed by adding different noises on input light curves, PELICAN has learned many noisy examples.


\subsection{Peak magnitudes}
The right panels of Fig. \ref{prop} show the accuracy as a function of the maximum value of peak magnitude from all bands. For the classification of DDF light curves, the accuracy decreases at low magnitudes above 26 due to the low number of examples in the training database in this magnitude range. However, PELICAN is robust at low magnitudes for the classification of WFD light curves. This robustness is due to the non-supervised learning during which PELICAN has learned a light curve representation at low magnitudes. Nevertheless, in this case, the accuracy decreases also at magnitudes below 23. This behaviour may be due to the mismatch between DDF light curves that composed the training database and WFD light curves from the test database. Indeed DDF light curves have, in average, brighter magnitudes than light curves in the test database. To reduce the mismatch between the training and test databases PELICAN performs a first non-supervised training on the test database. Nevertheless this step may cause a bias at bright magnitudes as PELICAN learned a representation of light curves at faint magnitudes from WFD survey.


\section{Summary and discussion }
We presented a deep learning architecture for the light curve classification, PELICAN. It performs several tasks to find the best feature representation space of light curves and classify them. In this work, we applied PELICAN to the analysis of supernovae light curves but it can be applied to the analysis of other variable and transient objects. Our model is able to reduce the problem of non-representativeness between the training and the test databases thanks to the development of two modules.
The first one uses a non-supervised autoencoder that benefits from light curves of the test set without knowing the labels in order to build a representative feature space. The second module optimizes a contrastive loss function adjusted to reduce the distance into the feature representation space between brighter and fainter objects of the same label. \\
PELICAN can also deal with the sparsity and the irregular sampling of light curves by integrating a sparsity parameter in the autoencoder module and performing an important data augmentation.\\
Our model reached best performance ever obtained for the Supernovae Photometric Classification Challenge with a non-representative training database, with an accuracy of 0.861 and an AUC of 0.937 against 0.713 and 0.855 respectively obtained by BDTS algorithm and SALT2 features  as shown in Lochner et al (2016). This kind of feature-based algorithms, does not permit to overcome the problem of representativeness. Indeed, even if feature used are relevant, they are not representative of the test database, as the spectroscopic follow-up is necessarily limited. Therefore this method, offers poor performance in a real scenario as we consider in this work, and have to be adapted. \\
In the context of LSST, it is important to confront PELICAN to the observational issues, in particular the uncertainties related to the two main programs of LSST which are the Wide Fast Deep and the Deep Drilling Fields surveys. In this work we addressed several points:
\begin{itemize}
    \item uncertainties related to the spectroscopic follow-up in DDF survey. A subsample of light curves should be spectroscopically confirmed in DDF survey but it might be very limited. PELICAN is able to reach good performance with small training database (2k light curves) for which it detects 87.4\% of SN Ia with a precision comparable to the spectroscopic one. 
    \item uncertainties related to the spectroscopic follow-up in WFD survey. It is not certain that a sample of light curves will be spectroscopically confirmed in WFD fields. So it is crucial that PELICAN could classify SN Ia observed on WFD survey, with a training composed only of DDF light curves. By considering a training database of 2k to 10k light curves, PELICAN is able to classify from 90.5\% to 97.3\% SN Ia with a precision higher than 98\%.  This result constitutes one of our major contribution as it opens the possibility of using SN Ia from WFD fields for cosmology studies.
\end{itemize}

We also found that PELICAN is robust against the SNR above 5 and magnitudes below 26 for the classification of DDF light curves. The accuracy of PELICAN is very stable until redshift 1.0, above this value the number of examples in the training database is not sufficient which explains the decrease at high redshifts. However this tendency is significantly reduced if the training database contains at least 5k light curves. In this case, the accuracy is higher than 90\% until 1.2 in redshift. \\ For the classification of WFD light curves the accuracy decreases at low redshifts and bright magnitudes, due to the mismatch between the training and test databases. Even if the step of non-supervised training on the test database reduces it, PELICAN learns more on low redshifts and faint magnitudes from the test database. It could be possible to reduce this bias by integrating spectroscopically confirmed light curves inside the training of the autoencodder but it should be done carefully as DDF light curves have to be transformed to look like WFD light curves to avoid the mismatch. Nevertheless, PELICAN remains robust at low SNR for the classification of WFD light curves. \\
PELICAN depends on the number of observations on the light curve as it takes only this information as input. Nevertheless the sparsity term in the loss of the autoencoder and the data augmentation, help to reduce the bias.  \\
A caveat for the tests done with simulation is the low number of Non Ia template available which may underestimate the proportion of non Ia that have similar lightcurves as Ia. This point will be adressed with more detailled simulators that will be available in the future.\\
Finally to complete validation of PELICAN, we tested it on real SDSS data. In this case there is a new mismatch that appears as we trained it on simulated data which do not well reproduce real SDSS data.  We demonstrated that an additional fraction of 10 \% of real light curves inside the training, allows to reach an accuracy of 86.8\%. This is a very encouraging result for the classification of supernovae light curves as the spectroscopically confirmed light curves can be completed by simulated ones to increase the size of the training database and so be less dependant on the costly spectroscopic follow-up. \\

The success of PELICAN under realistic conditions with a training step on a small and biased database and a testing stage on light curves with different sampling and more noisy measurements opens very promising perspectives for the classification of light curves of future large photometric surveys. Furthermore it constitutes, up to now, the most appropriate tool to overpass problems of representativeness on irregular and sparse data.

\section*{Acknowledgements}
This work has been carried out thanks to the support of the OCEVU Labex (ANR-11-LABX-0060). \\
We thank Rahul Biswas and Philippe Gris for useful discussions. \\
Funding for the creation and distribution of the SDSS Archive has been provided by the Alfred P. Sloan Foundation, the Participating Institutions, the National Aeronautics and Space Administration, the National Science Foundation, the US Department of Energy, the Japanese Monbukagakusho, and the Max Planck Society. The SDSS Web site is http://www.sdss.org/. The SDSS is managed by the Astrophysical Research Consortium (ARC) for the Participating Institutions. The Participating Institutions are the University of Chicago, Fermilab, the Institute for Advanced Study, the Japan Participation Group, The Johns Hopkins University, the Korean Scientist Group, Los Alamos National Laboratory, the Max-Planck-Institute for Astronomy (MPIA), the Max-Planck-Institute for Astrophysics (MPA), New Mexico State University, University of Pittsburgh, Princeton University, the United States Naval Observatory, and the University of Washington.

\bibliographystyle{aa}
\bibliography{biblio.bib} 


\end{document}